\UseRawInputEncoding

\documentclass[twocolumn]{aastex61}

\newcommand{\aips}{$\mathcal{AIPS}$\,}
\newcommand{\spirals}{S$\pi$RALS\,}
\newcommand{\kms}{km~s$^{-1}$}
\newcommand{\Fig}[1]{\hyperref[#1]{Figure~\ref*{#1}}}
\newcommand{\Tab}[1]{\hyperref[#1]{Table~\ref*{#1}}}
\newcommand{\Sec}[1]{\hyperref[#1]{Section~\ref*{#1}}}
\newcommand{\Eq}[1]{\hyperref[#1]{Equation~\ref*{#1}}}

\newcommand{\R}[1]{{\color{black}#1}}

\usepackage{savesym}

\usepackage{amsmath}
\restoresymbol{SIX}{tablenum}
\usepackage{hyperref}
\usepackage{booktabs}
\usepackage{comment}

\graphicspath{{figures/}}

\def\d    {\ifmmode {{\rlap{.}}^\circ}\else {${\rlap{.}}^\circ$}\fi}

\newcommand{\utas}{School of Natural Sciences, University of Tasmania, Private Bag 37, Hobart, Tasmania 7001, Australia}
\newcommand{\jive}{Joint Institute for VLBI ERIC, Oude Hoogeveensedijk 4, 7991PD Dwingeloo, Netherlands}
\newcommand{\cfa}{Center for Astrophysics $\vert$\ Harvard\ \&\ Smithsonian, Cambridge, MA\ 02138, USA}
\newcommand{\irasr}{Institute for Radio Astronomy and Space Research, Auckland University of Technology, 120 Mayoral Drive, Auckland 1010, New Zealand}
\newcommand{\icrar}{ICRAR, M468, The University of Western Australia, 35 Stirling Hwy, Crawley, Western Australia, 6009}
\newcommand{\mpifra}{Max-Planck-Institut für Radioastronomie, Auf dem Hügel 69, D-53121 Bonn, Germany}

\shorttitle{Inverse MultiView Calibration II}
\shortauthors{Hyland et al.}

\newcommand{\degs}{$^\circ$}

\begin{document}
	
	\title{Inverse MultiView II: Microarcsecond Trigonometric Parallaxes for Southern Hemisphere 6.7~GHz Methanol Masers G232.62+00.99 and G323.74--00.26 }

\author[0000-0002-4783-6679]{L. J. Hyland}\affil{\utas}
\author[0000-0001-7223-754X]{M. J. Reid}\affil{\cfa}
\author[0000-0002-5526-990X]{G. Orosz}\affil{\utas}\affil{\jive}
\author[0000-0002-1363-5457]{S. P. Ellingsen}\affil{\utas}
\author{S. D. Weston}\affil{\irasr}
\author[0000-0002-9571-8036]{J. Kumar}\affil{\utas}
\author[0000-0003-0392-3604]{R. Dodson}\affil{\icrar}
\author[0000-0003-4871-9535]{M.J. Roija}\affil{CSIRO Astronomy and Space Science, PO Box 1130, Bentley WA 6102, Australia}
\affil{\icrar}
\affil{Observatorio Astron\'omico Nacional (IGN), Alfonso XII, 3 y 5, 28014 Madrid, Spain}
\author[0000-0002-3297-9247]{W. J. Hankey}\affil{\utas}
\author[0000-0003-2806-3495]{P. M. Yates-Jones}\affil{\utas}
\author[0000-0002-1354-7510]{T. Natusch}\affil{\irasr}
\author[0000-0003-0186-5551]{S. Gulyaev}\affil{\irasr}
\author[0000-0001-6459-0669]{K. M. Menten}\affil{\mpifra}
\author[0000-0003-4468-761X]{A. Brunthaler}\affil{\mpifra}\

	
	\begin{abstract}
	We present the first results from the Southern Hemisphere Parallax Interferometric Radio Astrometry Legacy Survey (\spirals):  $10\mu$as-accurate parallaxes and proper motions for two southern hemisphere 6.7 GHz methanol masers obtained using the inverse MultiView calibration method. Using an array of radio telescopes in Australia and New Zealand, we measured the trigonometric parallax and proper motions for the masers associated with the star formation region G232.62+00.99 of $\pi = 0.610\pm0.011$~mas, $\mu_x=-2.266\pm0.021$~mas~y$^{-1}$ and $\mu_y=2.249\pm0.049$~mas~y$^{-1}$, which implies its distance to be $d=1.637\pm0.029$~kpc. These measurements represent an improvement in accuracy by more than a factor of 3 over the previous measurements obtained through Very Long Baseline Array observations of the 12~GHz methanol masers associated with this region. We also measure the trigonometric parallax and proper motion for G323.74--00.26 as $\pi = 0.364\pm0.009$~mas, $\mu_x=-3.239\pm0.025$~mas~y$^{-1}$ and $\mu_y=-3.976\pm0.039$~mas~y$^{-1}$, which implies a distance of $d=2.747\pm0.068$~kpc.  These are the most accurate measurements of trigonometric parallax obtained for 6.7~GHz class II methanol masers to date. We confirm that G232.62+00.99 is in the Local arm and find that G323.74--00.26 is in the Scutum-Centaurus arm. We also investigate the structure and internal dynamics of \R{both masers.}
	\end{abstract}
	
	
    \section{Introduction}
    \label{sec:intro}
    Measuring the trigonometric parallax (hereafter `parallax') and the proper motion of stars or star-forming regions that trace the motion of interstellar gas, is the best method to accurately determine the structure and kinematics of the Milky Way. Parallax measurements at radio frequencies have the advantage of not being obscured by dust, and can therefore probe much deeper into the disk of the Galaxy than those at optical frequencies.
    
    Very Long Baseline Interferometry (VLBI) has been demonstrated to be able to achieve parallax accuracies of $\pm10~\mu$as and therefore measure objects at a distance of 10~kpc with 10\% accuracy \citep{ReidHonma2014}. Thus far, this level of accuracy has been almost exclusively achieved at radio frequencies above 10~GHz and with homogeneous telescope arrays (e.g., the Very Long Baseline Array (VLBA)). At lower frequencies, uncompensated dispersive delays from the ionosphere can cause large systematic and direction-dependant errors \citep{Rioja2017,Reid2017,RiojaDodson2020}.

    MultiView \citep[hereafter `direct MV';][]{Rioja2017}, has been shown to give astrometric accuracies approaching values determined by the thermal noise. A new variation called inverse MultiView (iMV) has recently been developed that allows additional robust calibration of short-timescale quasi-random phases changes at the position of that target, for target-calibrator separations of up to 7\degs\ at 8.3~GHz \citep{Hyland2022}. 
    
	The Southern Hemisphere Parallax Interferometric Radio Astrometry Legacy Survey \citep[\spirals;][]{Hyland2021} is an extension of the Bar and Spiral Structure Legacy Survey \citep[BeSSeL;][]{Brunthaler2011,Reid2009_gal,Reid2014,Reid2019} and VLBI Exploration of Radio Astrometry \citep[VERA;][]{vera2020}, with the aim to obtain information on the structure of the Milky Way for those regions exclusively visible from the southern hemisphere. \spirals~targets 6.7~GHz methanol masers \citep{Menten1991}, which are known to exclusively trace high-mass star formation \citep{Minier2001,Ellingsen2007,Breen2013}. At these relatively low frequencies, the iMV approach can greatly improve the calibration of the dispersive delays due to the ionosphere.
	
	In this paper, we demonstrate the calibration capabilities of iMV by measuring 10$\mu$as-accurate parallaxes of two 6.7~GHz methanol masers. In \Sec{sec:sourceobs} we describe target and calibrator selection, and observations. In \Sec{sec:datareduction} we outline all data reduction and analysis, including iMV calibration and parallax fitting. \Sec{sec:results} presents the results and \Sec{sec:discussion} includes a discussion of our findings.

\section{Methods} \label{sec:sourceobs}
    \subsection{Source Selection and Observations}
    We selected two class II 6.7~GHz masers from the Methanol Multibeam Catalogue \citep[MMB; ][]{Caswell2010,Caswell2011,Green2012,Breen2015}, which our pilot observations revealed to have compact emission in at least one 2~kHz/0.087~\kms~velocity channel. The first was G232.62+00.99 \citep{MacLeod1992}, a maser associated with the \textsc{Hii} region RCW~7 \citep{DuboutCrillon1976}. As this target is at a declination of $-17^\circ$, it is visible from both the northern and southern hemispheres. Consequently, this maser region had an existing parallax measurement from the BeSSeL project \citep[at 12~GHz;][]{Reid2009_mas}. This allows a direct comparison of parallaxes measured by the BeSSeL survey and \spirals.

		\begin{figure}
			\centering
			\includegraphics[width=0.5\textwidth]{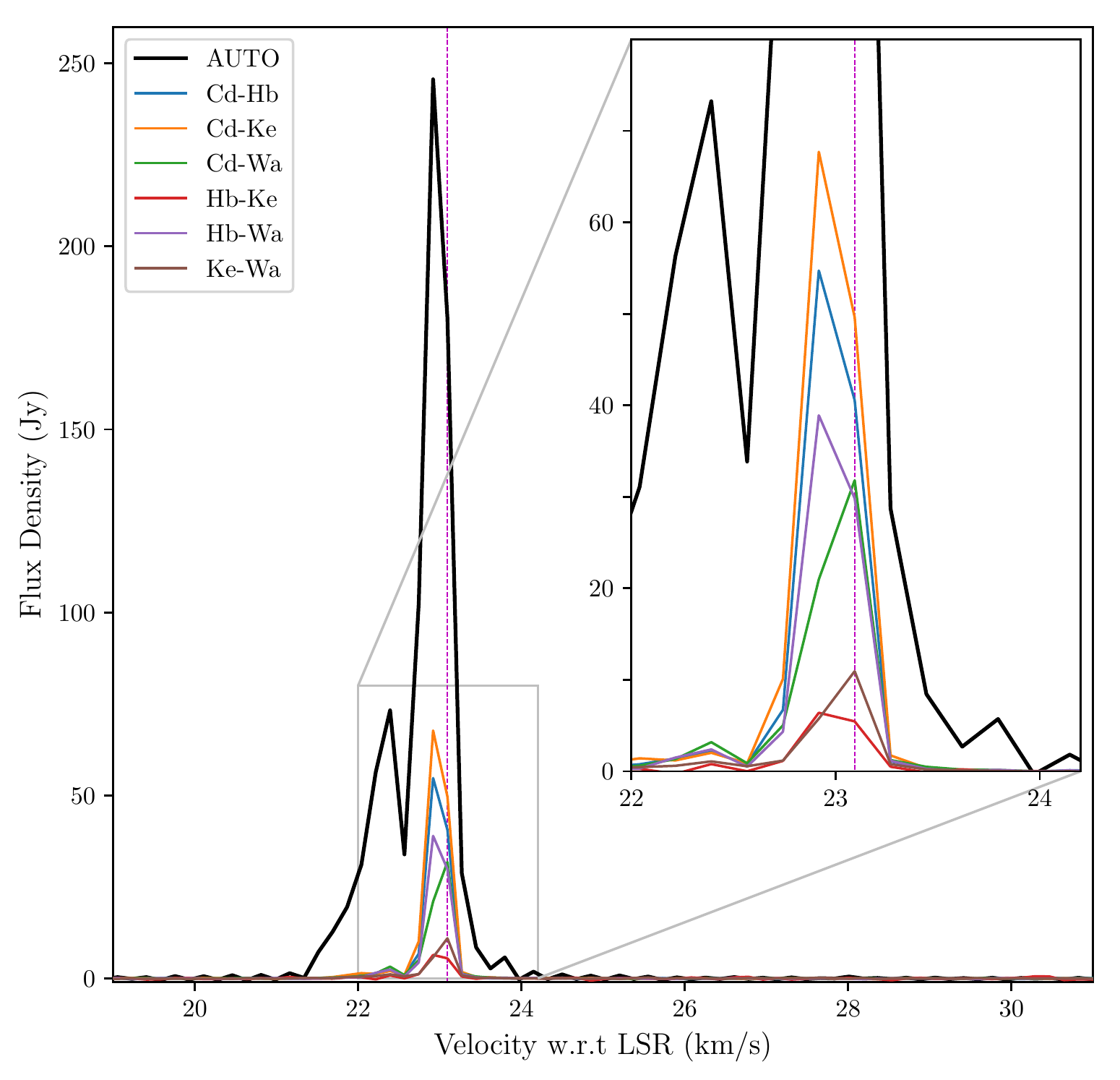}
			\caption{Auto- and cross-correlated spectra of G232.62+00.99 maser emission at epoch 6. Autocorrelated peak flux density (black line) is $\sim250$~Jy and Ke-Wa (the longest baseline) cross-correlated peak flux density is 10~Jy. The reference feature at 23.08~\kms~is indicated by the vertical magenta dashed line.}
			\label{fig:g232maserspectrum}
		\end{figure}
    
    The second maser was G323.74--00.26 \citep{MacLeod1992}, one of the strongest 6.7~GHz methanol masers known, which has been exhibiting a peak flux density over 3000~Jy \citep[while flares with flux density up to 5800 Jy have also been observed;][]{Goedhart2004}. This maser is at a declination of $-56^\circ$ and hence only visible to southern hemisphere instruments. There have been numerous studies of the 6.7~GHz methanol maser emission associated with this region
    \citep{Norris1993,Norris1998,Phillips1998,Walsh2002,Ellingsen2007,Vlemmings2011}.
    
	    \begin{figure}
		    \centering
		    \includegraphics[width=0.5\textwidth]{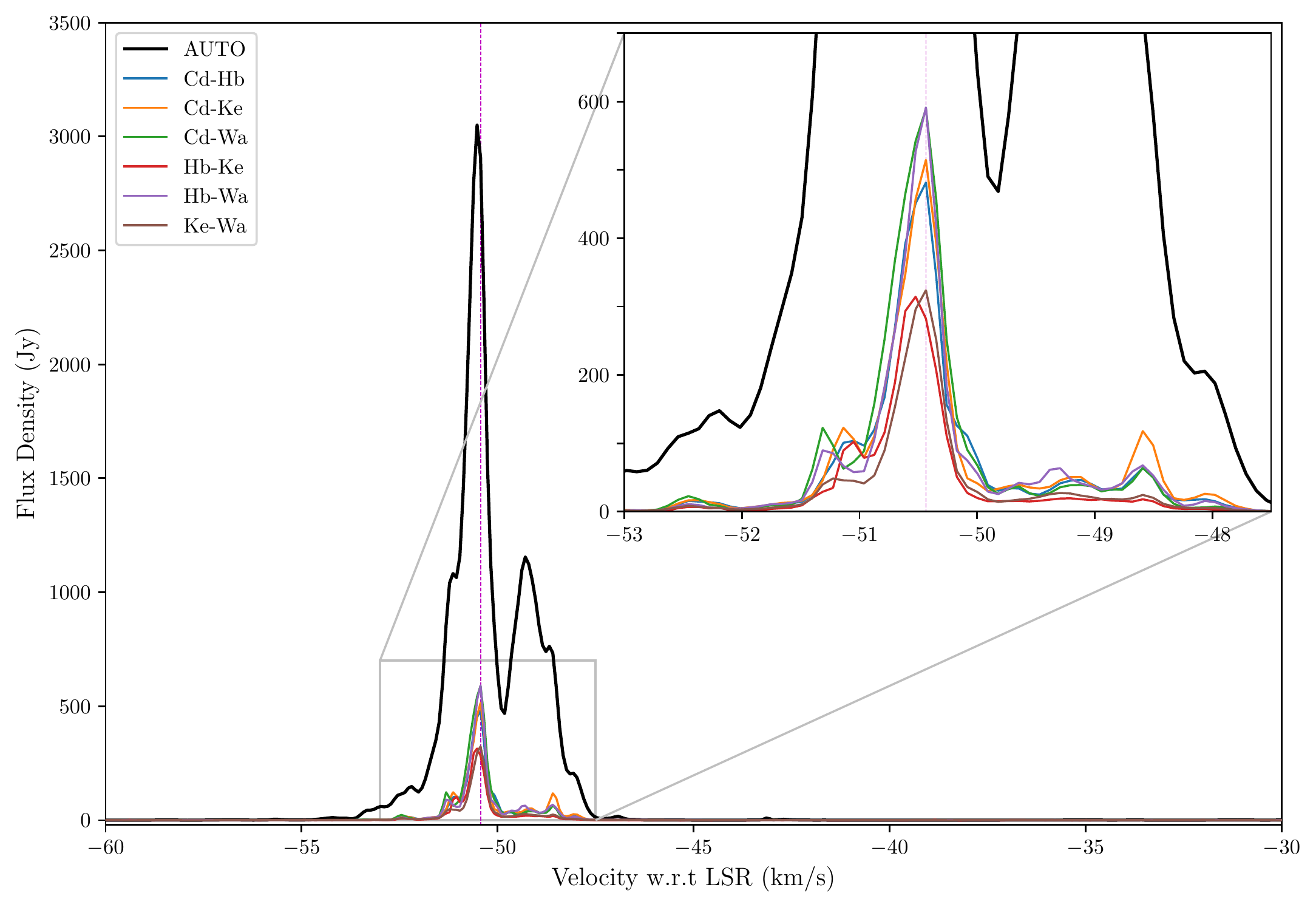}
		    \caption{Auto- and cross-correlated spectra of G323.74--00.26 maser emission at epoch 4. Autocorrelated peak flux density is $\sim3500$~Jy and Ke-Wa cross-correlated flux density (brown line) peaks as $300$~Jy at velocity $-50.52$~\kms\space (indicated by the vertical-dashed magenta line).}
		    \label{fig:g323maserspectrum}
	    \end{figure}    
     
     The spectra of G232.62+00.99 and G323.74--00.26 as detected by our array are shown in \Fig{fig:g232maserspectrum} and \Fig{fig:g323maserspectrum}, respectively.
     
    \citet{Hyland2022} showed that iMV with calibrators up to $\approx7$\degs\ separation from the target source can be successful at 8.3~GHz. At the lower frequency of 6.7~GHz, we chose calibrators separated by up to 5.5\degs. Allowing calibrators that are located up to this radius from a target provides many more bright and compact sources compared to the limitation that is usually applied for standard phase-referenced astrometric observations ($<2$\degs).

    Therefore for both of the target masers, we selected calibrators that met the criteria (in order of priority):
    \begin{itemize} 
    \item unresolved flux density $>100$~mJy;
    \item within 5.5\degs\ separation;
    \item uniform directional sky sampling (with the target near the center).
    \end{itemize}
    
    For G232.62+00.99, we chose four calibrators separated by between 1 and 4\degs\space from the target (\hyperref[tab:sources]{Table~\ref*{tab:sources}}), originally from the catalog of \citet{Petrov2019}. The fifth quasar at an angular separation of 5.3\degs\space(J0730--1141) was chosen as an electronic or ``manual-phase'' calibrator; however, its proximity to the target allowed  incorporation into the iMV cycle.
    
    For G323.74--00.26 we inferred the 6.7~GHz flux densities of calibrators from the available 8~GHz data \citep{Petrov2019}.  Using the selection criteria above, we chose the 6 calibrators listed in \Tab{tab:sources}. 
    Note, there were no good\footnote{Those exhibiting a flux density greater than 100~mJy, with positions known to better than 0.3~mas and/or with little to no extended structure.} calibrators known within 2\degs\ of G323.74--00.26, and therefore standard phase referencing (PR) would have been difficult and likely to produce poor astrometric results.

\begin{figure*}[ht]
			\centering
			\includegraphics[width=0.4\textwidth]{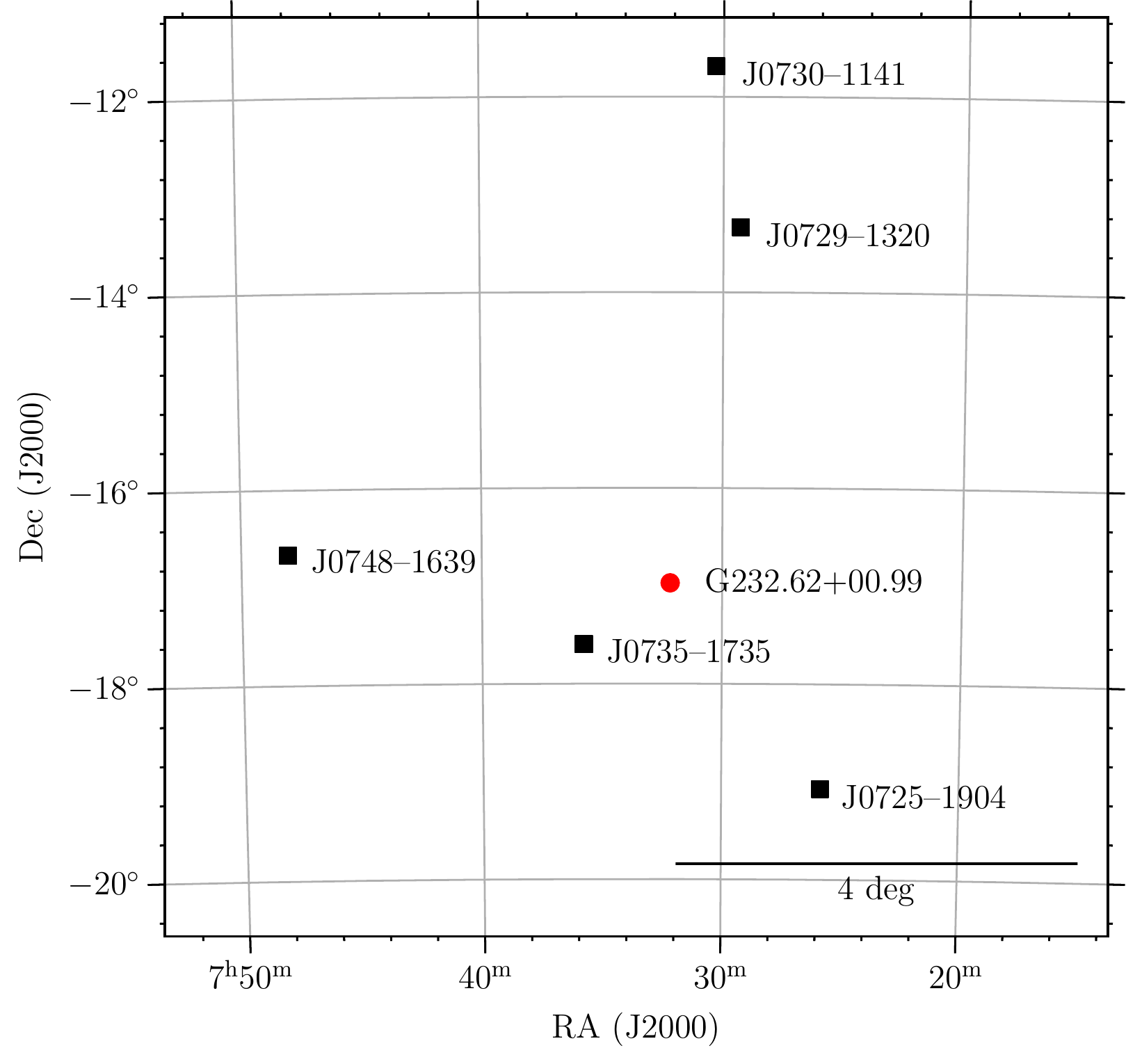}
			\includegraphics[width=0.4\textwidth]{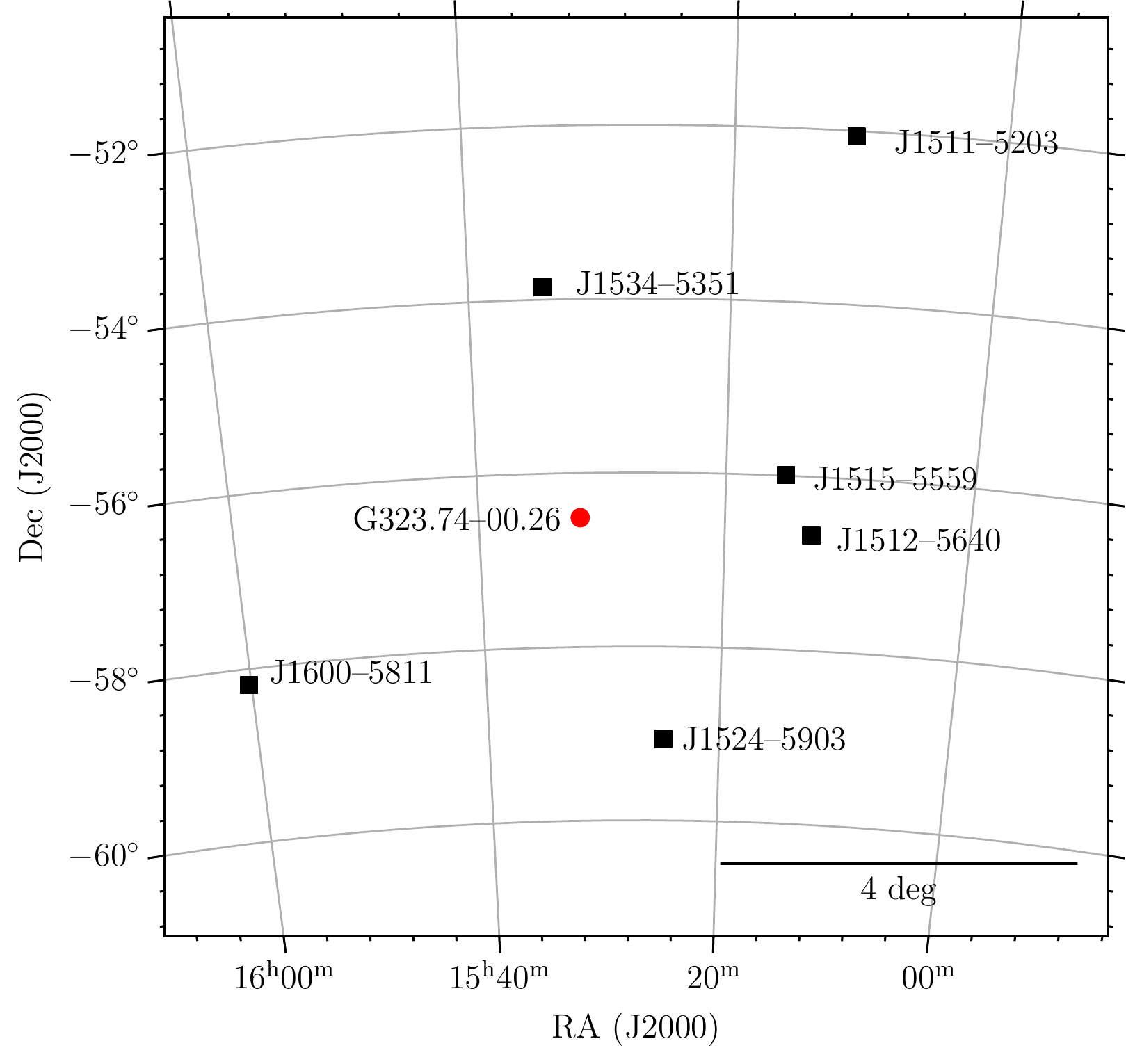}
			\caption{Sky distributions of target masers (red dot) and calibrators (black squares).  \textbf{Left:} Target maser G232.62+00.99 with 5 calibrator QSOs. The smallest calibrator separation is 1.06\degs\space(J0735--1735) and the largest is 5.30\degs\space(J1730--1141). \textbf{Right:} Target maser G323.740--0.263 with 6 calibrator QSOs. Smallest calibrator separation is 2.36\degs\space(J1515--5559) and largest is 5.37\degs\space(J1511--5203).}
			\label{fig:skyplots}
    	\end{figure*}

    The distribution of calibrators around the targets is shown in \Fig{fig:skyplots}, and \Tab{tab:sources} contains the calibrator positions and flux densities, listed in the order they were observed when nodding between the target and calibrator.
    
		\begin{table*}
			\small
			\centering
			\caption{Target maser and calibrator QSO \textit{correlated} positions, angular separations and flux densities. QSOs are ordered in the sequence that they were observed. \textbf{Columns:} Target maser and calibrator QSO (1-2) name, right ascension and declination position (3-4), angular separation in right ascension and declination (5-6), total separation (7), and flux density (8).}
			\begin{tabular}{llllcccc} \hline
				\multicolumn{2}{c}{\bf Source Name} & \multicolumn{1}{l}{\bf R.A.} & \multicolumn{1}{c}{\bf Dec.} & \multicolumn{3}{c}{\bf Separation} & \multicolumn{1}{c}{\bf Flux} \\
				\multicolumn{1}{c}{\bf Target} & \multicolumn{1}{c}{\bf Calibrator} & \multicolumn{1}{l}{(J2000)} & \multicolumn{1}{c}{(J2000)} & \multicolumn{1}{c}{$\Delta\alpha\cos\delta_T$} & \multicolumn{1}{c}{$\Delta\delta$} & \multicolumn{1}{c}{$\theta_\mathrm{sep}$} & \multicolumn{1}{c}{\bf Density} \\
				\multicolumn{1}{c}{\bf Maser} & \multicolumn{1}{c}{\bf QSOs} & \multicolumn{1}{l}{\bf $h$~~$m$~~$s$} & \multicolumn{1}{l}{\bf ~~$^\circ$~~$\prime$~~$\prime\prime$} & \multicolumn{1}{c}{(\degs)}  & \multicolumn{1}{c}{(\degs)}  & \multicolumn{1}{c}{(\degs)}  & \multicolumn{1}{c}{(Jy)} \\
				\hline
				G232.62+00.99 && 07~32~09.78     & --16~58~12.80    & & &                    & $\sim{1}0$\tablenote{\R{Correlated f}lux density of $+23.09$~\kms~channel at epoch 6 \R{on Ke-Wa baseline ($\sim4750$~km)}} \\
				& J0735--1735   & 07~35~45.812460 & --17~35~48.50242 &   0.86 & --0.62 & 1.06 & 0.10\\
				& J0725--1904   & 07~25~50.165557 & --19~04~19.07419 & --1.51 & --2.10 & 2.58 & 0.15 \\
				& J0729--1320   & 07~29~17.817692 & --13~20~02.27125 & --0.68 &   3.64 & 3.70 & 0.12 \\
				& J0748--1639   & 07~48~03.083813 & --16~39~50.25355 &   3.80 &   0.31 & 3.81 & 0.30 \\
				& J0730--1141   & 07~30~19.112473 & --11~41~12.60061 & --0.44 &   5.28 & 5.30 & 3.18 \\\hline
				G323.74--00.26 && 15~31~45.45     & --56~30~50.10    & & &                    & $\sim300$\tablenote{\R{Correlated f}lux density of $-50.52$~\kms~channel at epoch 4 \R{on Ke-Wa baseline ($\sim4750$~km)}} \\
				& J1534-5351    & 15~34~20.660723 & --53~51~13.42272 &   0.36 &   2.66 & 2.68 & 0.13\tablenote{\R{Catalogued u}nresolved flux density at 8.4~GHz.} \\
                & J1600--5811    & 16~00~12.377460 & --58~11~02.96855 &   3.92 & --1.67 & 4.18 & 0.32$^\mathrm{c}$ \\
                & J1524--5903    & 15~24~51.122912 & --59~03~39.71702 & --0.95 & --2.55 & 2.71 & 0.06$^\mathrm{c}$ \\
                & J1512--5640    & 15~12~55.819395 & --56~40~30.64300 & --2.60 & --0.16 & 2.60 & 0.20$^\mathrm{c}$ \\
                & J1515--5559    & 15~15~12.672909 & --55~59~32.83823 & --2.28 &   0.52 & 2.36 & 0.26$^\mathrm{c}$ \\
                & J1511--5203    & 15~11~08.926191 & --52~03~47.25032 & --2.84 &   4.45 & 5.37 & 0.05$^\mathrm{c}$ \\ \bottomrule
			\end{tabular}
			\label{tab:sources}
		\end{table*}
		
    The structure of an individual observation session of nine hours was almost identical to that used by  \cite{Hyland2022}, with MultiView blocks bracketed by ``geodetic-like'' calibration blocks \citep{Honma2007,Reid2009_mas,ReidHonma2014} and scans on bright compact calibrators. We observed seven epochs over a period of 1.5~years for G232.62+00.99 and seven epochs spanning 1~year for G323.74--00.26. The dates of observations were chosen to sample near the extremes of the parallax oscillation in right ascension (R.A.) in order to optimize the accuracy to which the observations can measure the parallax. We refer to these maxima as ``parallax seasons''.

    \subsection{Array, Frequency and Correlation} \label{sec:arrfreq}
\begin{figure}
    	\centering
    	\includegraphics[width=0.5\textwidth]{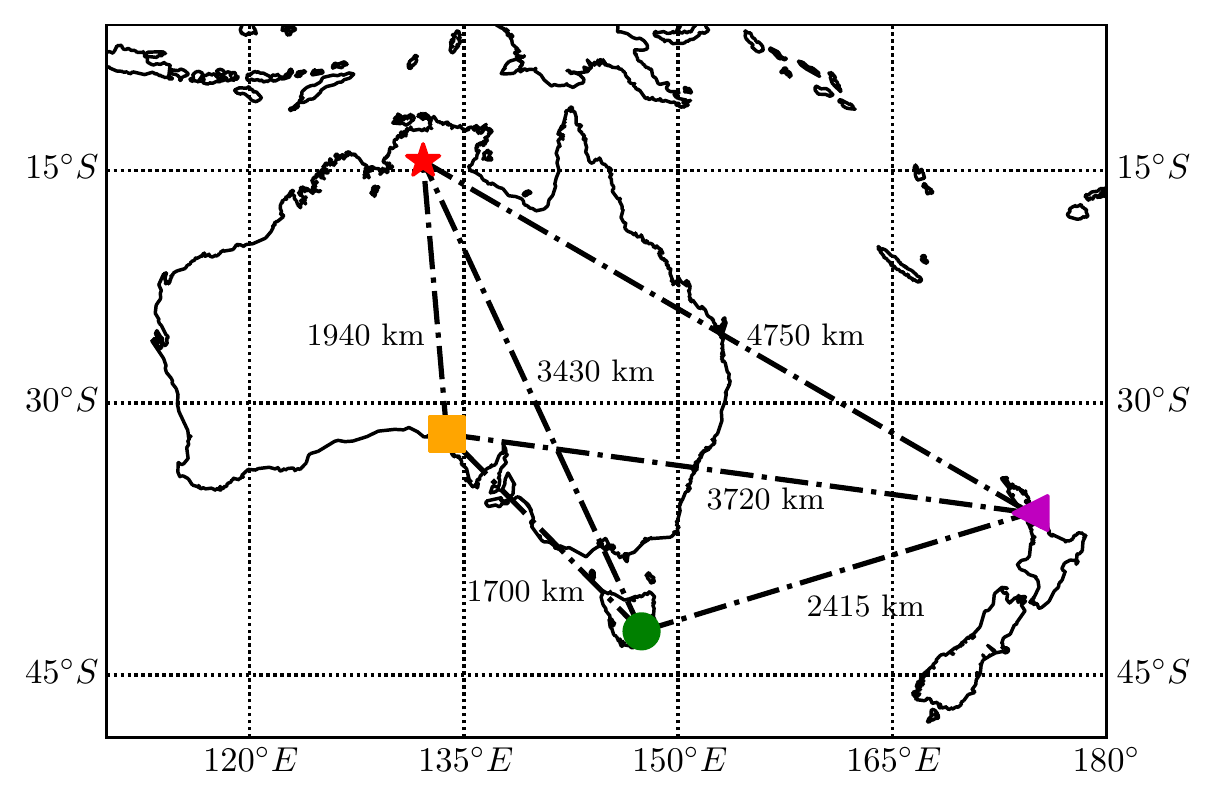}
    	\caption{VLBI array used for these observations, with the University of Tasmania telescopes Ceduna~30m (orange square), Hobart~12m (green circle), Katherine~12m (red star) and Auckland University of Technology telescope Warkworth~30m (purple triangle). Baselines are indicated with dash-dotted lines and baseline lengths are given next to corresponding baselines.}
    	\label{fig:array}
    \end{figure}

    The array used for these observations is shown in \hyperref[fig:array]{Figure~\ref*{fig:array}}, comprising the University of Tasmania telescopes Ceduna~30m \citep[Cd; ][]{McCulloch2005}, Hobart~12m (Hb) and Katherine~12m \citep[Ke; ][]{Lovell2013}, and the Auckland University of Technology telescope Warkworth~30m \citep[Wa;][]{Woodburn2015}. This array has a maximum baseline length of 4750~km.
    
    We recorded data at 1024~{\R M}bps in dual polarisation, Nyquist sampled at 2-bits per sample, over the frequency range 6580--6708~GHz. The Ceduna~30m and Warkworth~30m antennas recorded right and left ($\mathcal{R,L}$) circular polarisations while Hobart~12m and Katherine~12m recorded horizontal and vertical ($\mathcal{H,V}$) linear polarizations. 
    
    Baseband data were correlated using DiFX-2 \citep{Deller2011} in two passes: all sources were correlated over the full recorded frequency range with 0.5~MHz frequency channels, and the iMV blocks were correlated in a 4~MHz ``zoom'' band approximately centered on the frequency of the maser emission. The zoom band data for G232.62+00.99 were correlated to give frequency channels of 4~kHz, corresponding to velocity channels of 0.174~km~s$^{-1}$; for G323.74--00.26 we used frequency channels of 2~kHz giving velocity channels of 0.087~\kms.
    
    In order to create valid FITS files from our mixed polarisation (i.e., $\mathcal{HR}$, $\mathcal{VR}$, $\mathcal{HL}$, $\mathcal{VL}$, $\mathcal{RR}$, $\mathcal{LL}$, $\mathcal{RL}$, $\mathcal{LR}$) data (in \textit{difx2fits}), we treated Hobart~12m and Katherine~12m $\mathcal{H}$ as $\mathcal{R}$ and $\mathcal{V}$ as $\mathcal{L}$ in our correlation.  This approach was required as we were unsuccessful at converting the mixed polarisation products to a pure circular basis, owing to a lack of understanding of the complex apparent feed rotation characteristics of the Warkworth 30m (see \Sec{sec:wapangle}). This \R{would} reduce the amplitudes of mixed polarization products by $\sqrt{2}$ (for an unpolarized source). Methanol masers at 6.7~GHz are known to exhibit low linear (typically 1.0–-2.5\%, max 17\%) and circular polarization \citep[typically 0.5--0.75\%, max 6\%;][]{Surcis2022}. \R{The visibility amplitudes of the mixed products on both masers does not differ more than 20\%, where we attribute most of this to a lack of exact amplitude and/or polarization calibration \citep[e.g.,][]{Dodson2008}}. In either case, we did not see any \R{obvious} adverse effects on our astrometric accuracy. 
    
\section{Data Reduction and Analysis} \label{sec:datareduction}
\subsection{Preliminary reduction}
    We calibrated correlated data in a similar manner to \cite{Hyland2022}, with a few changes to account for dual-polarization/spectral line data, and that the targets may have significant positional changes over the year(s). Briefly, using \aips/ParselTongue \citep{Greisen1990,Greisen2003,Kettenis2006} the additional steps to calibrate the data were as follows:
    \begin{enumerate}
    	\item Apparent feed rotation corrections were applied to the dual polarisation data with task CLCOR/PANG (see \Sec{sec:wapangle} for a detailed discussion of the complex correction needed for the Warkworth~30m antenna).
        \item Source position shifts (to improve relative positions among the sources) were applied with task CLCAL/ANTP.
    	\item Tasks SETJY/CVEL were used to correct for the Earth's (rotation and orbital) Doppler shift, ensuring that the maser spectra were aligned in frequency during and  across epochs.
    	\item A maser channel was selected to be the phase reference and the task FRING was used to solve for the phase and rate on that single channel. The criteria for channel selection was maximum flux density on the long Katherine~12m--Warthworth~30m baseline. 
    	\item The visibilities for the continuum sources were averaged in frequency using the task SPLIT, then the phases and rates from the maser reference channel were applied.
    	\item The calibrators were imaged using the IMAGR task.  The position offset was determined for each calibrator and the weighted offsets of all calibrators were assumed to reflect a position error in the reference maser spot. The maser position corrections were measured for each observing season $(x_T,y_T)$. Individual calibrator source offsets were used to refine their positions {\it relative} to the new maser position at one epoch.  All position offsets were applied to each source in Step 2.  \textit{Important note:} The final calibrator positions were required to be consistent for all epochs. 
        
        \item The calibrators were averaged in frequency using the task SPLAT and the phase was measured with the CALIB task. The solutions were output and used for iMV fitting (described in \Sec{subsec:iMV})
    \end{enumerate}

    \subsection{Warkworth~30m Apparent Feed Rotation Correction} \label{sec:wapangle}
    The Warkworth~30m antenna is a Nasmyth wheel-on-track antenna with a beam waveguide design \citep{Petrov2015}. The physical feed does not rotate when the antenna moves and a system of four mirrors directs the beam into the receiver; i.e., a Beam Wave-Guide system, which is not uncommon for converted telecom antennas (Warkworth NZ, Yamaguchi JP, Nkutunse GH, etc.). At the time of analysis, the standard \aips task CLCOR/PANG did not include corrections for this type of focus. In order to combine the dual polarisation data 
    we needed to correct for the phase introduced as the antenna moves in azimuth and elevation.
    
    We found the phase correction, $\varphi$, that accounts for the apparent feed rotation is:
    \begin{equation}
    \varphi = -q - \mathcal{A} + \mathcal{E}
    \end{equation} where $q$ is the parallactic angle, $\mathcal{A}$ is the azimuth angle (measured North through East), and $\mathcal{E}$ is the elevation angle. Subtracting  $\varphi$ from the RCP signal ($\mathcal{R}$) and adding it to the LCP signal ($\mathcal{L}$) phase corrects the visibility data for the apparent feed rotation, allowing the $\mathcal{R}$ and $\mathcal{L}$ data to be averaged before fringe fitting on the maser and increasing the S/N by a factor of $\sqrt{2}$. The feed correction has subsequently been added to the CLCOR task and the technical details are described in \citet{beamwaveguide}.
    
    \subsection{iMV Calibration}\label{subsec:iMV}
    Given the existence of occasional phase wraps in the 8.3~GHz experiments described by \citet{Hyland2022}, we expected a similar or greater number to be present at 6.7 GHz. The residual path delay, $\Delta\tau$, for a dispersive medium like the ionosphere scales with frequency, $\nu$, as $\nu^{-2}$.  Interferometer phase, $\phi$, is given by $\phi = \Delta\tau~\nu$, and thus the effect on phase scales as $\nu^{-1}$.
    Therefore, scaling from 8.3~GHz to 6.7~GHz should lead to a 20\% increase in phase shifts (assuming the same value for the residual total electron content TEC). This implies there is likely to be an increased number of phase wraps in our 6.7~GHz data compared to those seen by \citet{Hyland2022}.
    
    In order to unwrap the phases, we took the minimum difference of phase between consecutive scans on the same calibrator when adding trial values of 360\degs, 0\degs, and $-360$\degs. Additionally, all phases were minimized relative to a common time at the center of the track, where the delay errors due to residual tropospheric and ionospheric errors are expected to be at a minimum. 
    
    Once unwrapped, the phase data on each scan was fit with the least squares method to a model for a 2D plane (see Equation 5 from \cite{Hyland2022}) and the interpolated phase at the origin was subtracted from the maser visibility data using the task CLCAL.
    
    The maser reference channel was then imaged using the \aips~task IMAGR, and the brightness distribution was fitted with a Gaussian model using the task JMFIT in order to measure the astrometric offsets ($x_m,y_m$) from the original phase center in \Tab{tab:shiftsoffsets}.
 
	\begin{figure*}[ht]
		\centering
		\includegraphics[width=0.85\textwidth]{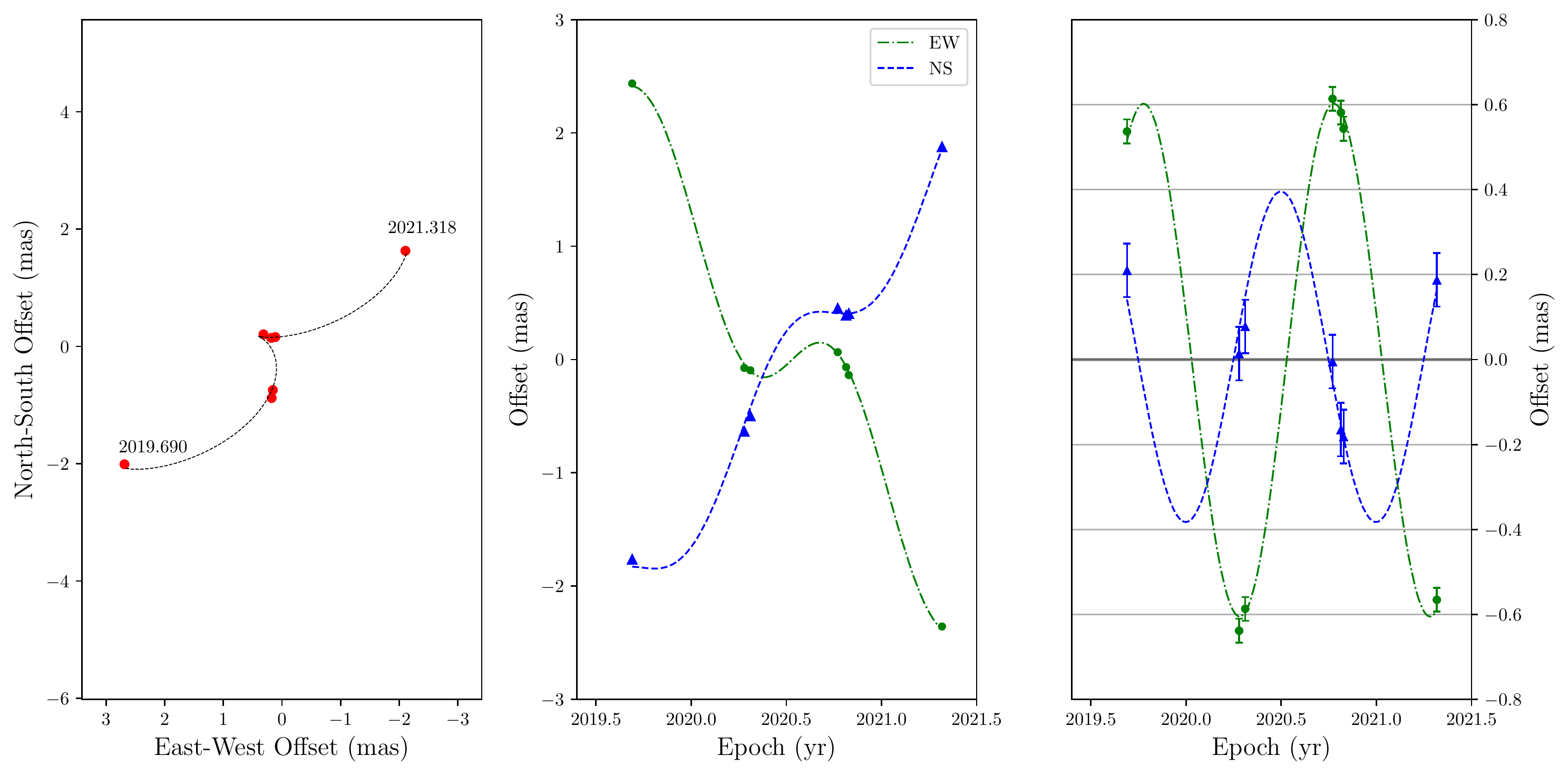}
		\caption{Parallax and proper motion modelling of G232.62+00.99 velocity channel $v=23.08$~\kms. \textbf{Left:} Total sky position change over the full observation period. \textbf{Middle:} Decomposition of sky motion into East-West (green dot-dashed) and North-South (blue dashed) motion over time. \textbf{Right:} Proper--motion subtracted sky motion over time. Error bars include 1$\sigma$ error floors for each coordinate of $28\mu$as and $62\mu$as for the EW and NS directions respectively.}
	\label{fig:g232par}
	\end{figure*}
 
    \subsection{Position shifting and parallax fitting}
    In order to minimize phase wraps (see the previous section), the (moving) maser position was updated at each observing season.  In order to put the measured positions back into a stationary reference frame, one must undo these shifts before fitting the parallax and proper motion.
    To achieve this, we first \R{chose a reference position shift from the} correlated position: ${\widetilde{x}_T,\widetilde{y}_T}$. We then calculated the offset from this reference position at each epoch (i.e., $x_T-\widetilde{x}_T$) and added it to the measured maser offset from the synthesized images ($x_m,y_m$). This gave the total offset from the reference position over time ($x_\mathrm{tot},y_\mathrm{tot}$). These values are given in \Tab{tab:shiftsoffsets}.
    
\begin{figure*}[ht]
		\centering
		\includegraphics[width=0.85\textwidth]{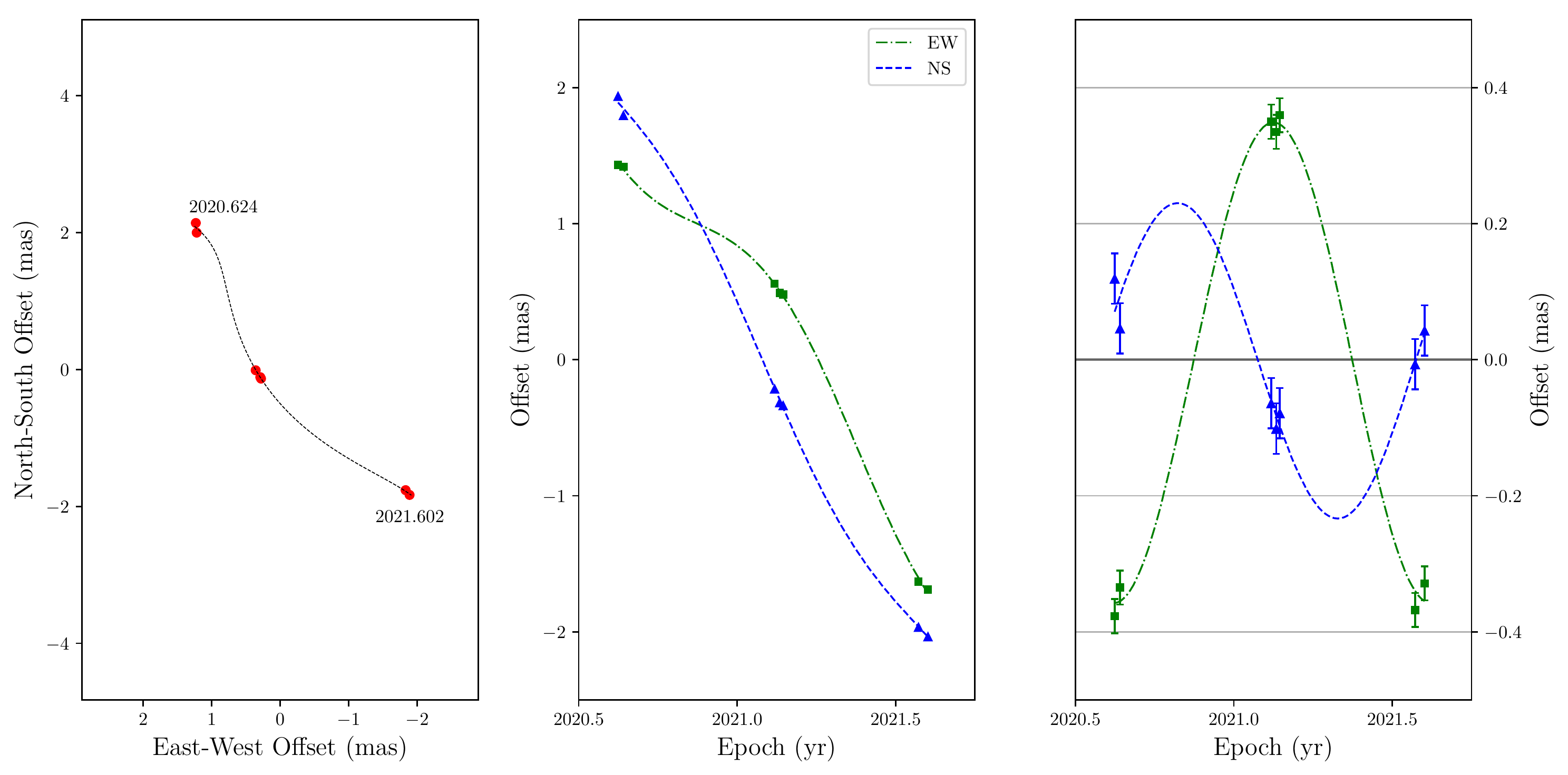}
		\caption{Parallax and proper motion modelling of G323.74--00.26 velocity channel $v=-50.52$~\kms. \textbf{Left:} Total sky position change over the full observation period. \textbf{Middle:} Decomposition of sky motion into East--West (green dot-dashed) and North--South (blue dashed) motion over time.  \textbf{Right:} Parallax motion (i.e., proper--motion subtracted) over time.  Error bars show independent 1$\sigma$ error floors for each direction as $25\mu$as and $39\mu$as for the EW and NS directions respectively.}
	\label{fig:g323par}
	\end{figure*}

    We fit the ($x_\mathrm{tot},y_\mathrm{tot}$) data using the model of \Eq{eq:model_parallax} in the Appendix with variance-weighted least-squares to determine the parallax ($\pi$), and the proper motions ($\mu_x,\mu_y$).
    Since astrometric uncertainty is usually dominated by systematic error, whose magnitude
    is not known {\em a priori}, we added ``error floors'' to the $x$ and $y$ data in quadrature. We independently varied these error floors to achieve a chi-squared per degree of freedom of unity for each coordinate.  This approach  is widely used in maser astrometry and is considered the most reliable method for estimating the uncertainties in $\pi,\mu_x,\mu_y$ \citep{Reid2009_mas}.
    The models as fit to the astrometric data for each target are shown in \Fig{fig:g232par} and \Fig{fig:g323par}.

    \begin{table*}[t]
        \caption{Epochs of VLBI observations and astrometric positions.}
        \centering
        \begin{tabular}{rcccc|cc|cc|cc} \toprule
		\multicolumn{1}{c}{\bf Target} &
		\multicolumn{1}{c}{\bf Epoch}  &
		\multicolumn{1}{c}{\bf Date}  &
		\multicolumn{2}{c}{\bf Target Shift}   &
		\multicolumn{2}{c}{\bf w.r.t Ref. pos}   &
		\multicolumn{2}{c}{\bf Meas. Offset}   &
		\multicolumn{2}{c}{\bf Total Offset}   \\
		\multicolumn{3}{c}{} &
		\multicolumn{1}{c}{${x_T}$}&\multicolumn{1}{c}{${y_T}$}&
		\multicolumn{1}{c}{${x_T-\widetilde{x}_T}$}&\multicolumn{1}{c}{${y_T-\widetilde{y}_T}$}&
		\multicolumn{1}{c}{${x_m}$}&\multicolumn{1}{c}{${y_m}$}&
		\multicolumn{1}{c}{${x_\mathrm{tot}}$}&\multicolumn{1}{c}{${y_\mathrm{tot}}$} \\
		\multicolumn{3}{c}{}&
		\multicolumn{1}{c}{(mas)}& 
		\multicolumn{1}{c}{(mas)}&
		\multicolumn{1}{c}{(mas)}& 
		\multicolumn{1}{c}{(mas)}&
		\multicolumn{1}{c}{(mas)}& 
		\multicolumn{1}{c}{(mas)}& 		
		\multicolumn{1}{c}{(mas)}& 
		\multicolumn{1}{c}{(mas)}\\\hline
  G232.62+00.99 & 1 & 2019~Sep~09 & 20.82 & 242.083 & 2.7 &-2.2 &-0.028 & 0.177 & 2.692 &-2.023\\ 
                & 2 & 2020~Apr~12 & 18.42 & 243.583 & 0.3 &-0.7 &-0.146 &-0.203 & 0.154 &-0.903\\ 
                & 3 & 2020~Apr~24 & 18.42 & 243.583 & 0.3 &-0.7 &-0.212 &-0.263 & 0.088 &-0.963\\ 
                & 4 & 2020~Oct~09 & 18.22 & 244.383 & 0.1 & 0.1 & 0.208 & 0.110 & 0.308 & 0.210\\ 
                & 5 & 2020~Oct~25 & 18.22 & 244.383 & 0.1 & 0.1 &-0.009 & 0.093 & 0.091 & 0.193\\ 
                & 6 & 2020~Oct~30 & 18.22 & 244.383 & 0.1 & 0.1 &-0.043 & 0.063 & 0.057 & 0.163\\ 
                & 7 & 2021~Apr~26 & 16.22 & 245.883 &-1.9 & 1.6 &-0.203 &-0.017 &-2.103 & 1.583\\
                & & \textit{Ref. pos} ($\boldsymbol{\widetilde{x}_T,\widetilde{y}_T}$): &  18.12 & 244.283  &&&&&& \\  
                \hline 
 G323.74--00.26 & 1 & 2020~Aug~15 &-187.1 &-177.7 & 1.5 & 1.9 &-0.268 &  0.239 & 1.232 & 2.139 \\ 
                & 2 & 2020~Aug~21 &-187.1 &-177.7 & 1.5 & 1.9 &-0.281 &  0.098 & 1.219 & 1.998 \\ 
                & 3 & 2021~Feb~12 &-188.3 &-179.6 & 0.3 & 0.0 & 0.058 &  0.012 & 0.358 &-0.012 \\ 
                & 4 & 2021~Feb~18 &-188.3 &-179.6 & 0.3 & 0.0 &-0.009 & -0.143 & 0.291 &-0.143 \\ 
                & 5 & 2021~Feb~22 &-188.3 &-179.6 & 0.3 & 0.0 &-0.020 & -0.114 & 0.280 &-0.114 \\ 
                & 6 & 2021~Jul~28 &-190.4 &-181.4 &-1.8 &-1.8 &-0.031 &  0.037 &-1.831 &-1.763 \\ 
                & 7 & 2021~Aug~08 &-190.4 &-181.4 &-1.8 &-1.8 &-0.089 & -0.033 &-1.889 &-1.833 \\ 
                & & \textit{Ref. pos} ($\boldsymbol{\widetilde{x}_T,\widetilde{y}_T}$): &  -188.6 &-179.6  &&&&&& \\ 
                \hline\bottomrule
        \end{tabular}
        \label{tab:shiftsoffsets}
    \end{table*}

\section{Results } \label{sec:results}
	\begin{table*}
		\centering
		\caption{Trigonometric parallaxes and proper motions determined from iPR from the target maser, and iMV.}
		\begin{tabular}{lllcccc} \hline
			\multicolumn{2}{c}{\bf Source} & \multicolumn{1}{c}{$\boldsymbol{\theta_\mathrm{sep}}$} & \multicolumn{1}{c}{$\boldsymbol{\pi}$} & \multicolumn{1}{c}{$\boldsymbol{\mu_x}$} & \multicolumn{1}{c}{$\boldsymbol{\mu_y}$} & \multicolumn{1}{c}{\bf Num.} \\ 
			\multicolumn{1}{l}{\bf Target} & \multicolumn{1}{l}{\bf Background} & \multicolumn{1}{c}{(\degs)} &\multicolumn{1}{c}{(mas)}& \multicolumn{1}{c}{(mas~y$^{-1}$)} & \multicolumn{1}{c}{(mas~y$^{-1}$)} & {\bf Epochs}\\
			\midrule
		 G232.62+00.99 &J0735--1735 & 1.06 &$0.523\pm0.055$ & $-2.367\pm0.100$ & $2.353\pm0.100$ & 6\\ 
			     &J0725--1904 & 2.58 &$0.634\pm0.053$ & $-2.221\pm0.098$ & $2.155\pm0.204$ & 7 \\ 
			     &J0729--1320 & 3.70 &$0.636\pm0.048$ & $-2.175\pm0.085$ & $2.533\pm0.128$ & 6\\ 
			     &J0748--1639 & 3.81 &$0.433\pm0.024$ & $-2.252\pm0.043$ & $2.958\pm0.142$ & 6\\ 
			     &J0730--1141\tablenote{iPR from G232.62+00.99 to QSO J0730--1141 did not produce coherent \R{synthesized images} at any epoch.} & 5.30 & - &-&-&-\\\smallskip
			&\textit{iMV w/ all}&& $0.610\pm0.011$ & $-2.266\pm0.021$ & $2.249\pm0.049$ & 7  \\
            &\textit{inc. int. mot.} & && $-2.266\pm0.330$ & $2.249\pm0.330$ &\\\midrule
		G323.74--00.26 & J1515--5559 & 2.36  & $0.359\pm0.027$ & $-3.25\pm0.08$ & $-4.01\pm0.10$  &7 \\
				& J1512--5640 & 2.60  & $0.374\pm0.032$ & $-3.23\pm0.09$ & $-3.83\pm0.25$ & 7 \\
	            & J1534--5351 & 2.68  & $0.437\pm0.028$ & $-3.20\pm0.08$ & $-3.89\pm0.15$ & 7 \\
				& J1524--5903\tablenote{iPR from G323.74--00.26 to QSO J1524--5903 only produced coherent at three out of seven epochs.} & 2.71 & - & - & - & 3   \\
				& J1600--5811 & 4.18  & $0.294\pm0.051$ & $-3.06\pm0.14$ & $-3.72\pm0.19$ & 7 \\
				& J1511--5203 & 5.37  & $0.484\pm0.059$ & $-3.57\pm0.18$ & $-3.71\pm0.25$ & 6 \\\smallskip
		&\textit{iMV w/ all}  & & $0.364\pm0.009$ &$-3.239\pm0.025$ & $-3.976\pm0.039$  & 7\\
		&\textit{inc. int. mot.} & &&$-3.239\pm0.400$ & $-4.174\pm0.400$  \\\bottomrule
		\end{tabular}
	    \tablecomments{iMV results attained using data from all QSOs. {\it Inc. int. mot.} abbreviates including internal motions.}
		\label{tab:parallax}
	\end{table*}
	
	Using iMV, we measure the parallax and proper motion of the 6.7~GHz emission in G232.62+00.99 to be $\pi=0.610\pm0.011$~mas, $\mu_x=-2.266\pm0.021$~mas~y$^{-1}$, and $\mu_y=2.249\pm0.049$~mas~y$^{-1}$.  
	This yields a distance of $d=1.639\pm0.030$~kpc (i.e., parallax inversion $d=1/\pi$ with symmetric errors $\sigma_d=\sigma_\pi/\pi^2$).
	For G323.74--00.26, we measure a parallax of $\pi=0.364\pm0.009$~mas and proper motion of $\mu_x=-3.239\pm0.025$ and $\mu_y=-3.976\pm0.039$~mas~y$^{-1}$, yielding a distance of $d=2.747\pm0.068$~kpc.

    In order to evaluate the astrometric improvement of iMV compared to standard (inverse)
	phase referencing using a single-calibrator, we also estimated parallaxes relative to each quasar in each cluster.   We find that the parallaxes based on individual calibrators in the G232.62+00.99 cluster range from 0.433 to 0.636~mas and in the G323.74--00.26 cluster range from 0.294 to 0.484~mas.  These results are consistent with systematic parallax 
	shifts of magnitude $\sim0.05$ mas per degree of calibrator angular offset found by \citet{Reid2017}. All of the parallax and proper motion results are tabulated in \Tab{tab:parallax}.

\section{Discussion}\label{sec:discussion}
  \subsection{G232.62+00.99}
   	The parallax and proper motion of the 12~GHz methanol emission in G232.62+1.0 were measured with the VLBA between October 2005 and March 2007 to be $\pi=0.596\pm0.035$~mas, $\mu_x =-2.17\pm0.06$~mas~y$^{-1}$, $\mu_y = 2.09\pm0.46$~mas~y$^{-1}$ \citep{Reid2009_mas}. Compared to this previous measurement, the parallax and proper motions of the 6.7~GHz methanol and 12~GHz methanol masers agree within the quoted uncertainties, with the obvious difference being that the new estimate is 3-times more accurate for $\pi$ and $\mu_x$ and an order of magnitude more accurate for $\mu_y$. 
   	
	It should be noted that the previous measurement was subject to issues that limited the performance of the VLBA: namely the source was observed at very low elevations and, the 12~GHz emission was resolved and only the inner five VLBA antennas were used, limiting the maximum baseline length to 1500~km (compared to the maximum baseline we use of 4750~km). Accounting for the latter by simply dividing the previous parallax measurement uncertainty by $\sim3$ reduces it to $\pm12~\mu$as. This indicates that we were able to successfully calibrate the delays (primarily the ionosphere) at 6.7~GHz to at least the same levels as could be achieved at 12~GHz, if not better.
    
    The previous southern hemisphere 6.7~GHz methanol maser parallaxes were measured by \citet{Krishnan2015,Krishnan2017} on the Long Baseline Array -- another southern hemisphere VLBI array that has common telescopes with the array that we used. These measurements were plagued by uncompensated dispersive delays, leading to parallaxes with accuracy between $50-110~\mu$as. Compared to the $\sim10~\mu$as parallaxes we have measured, we can see there has been a marked improvement owing to MV techniques.

\begin{figure}[ht]
			\centering
			\includegraphics[width=0.44\textwidth,trim={0 1.5cm 0 0}]{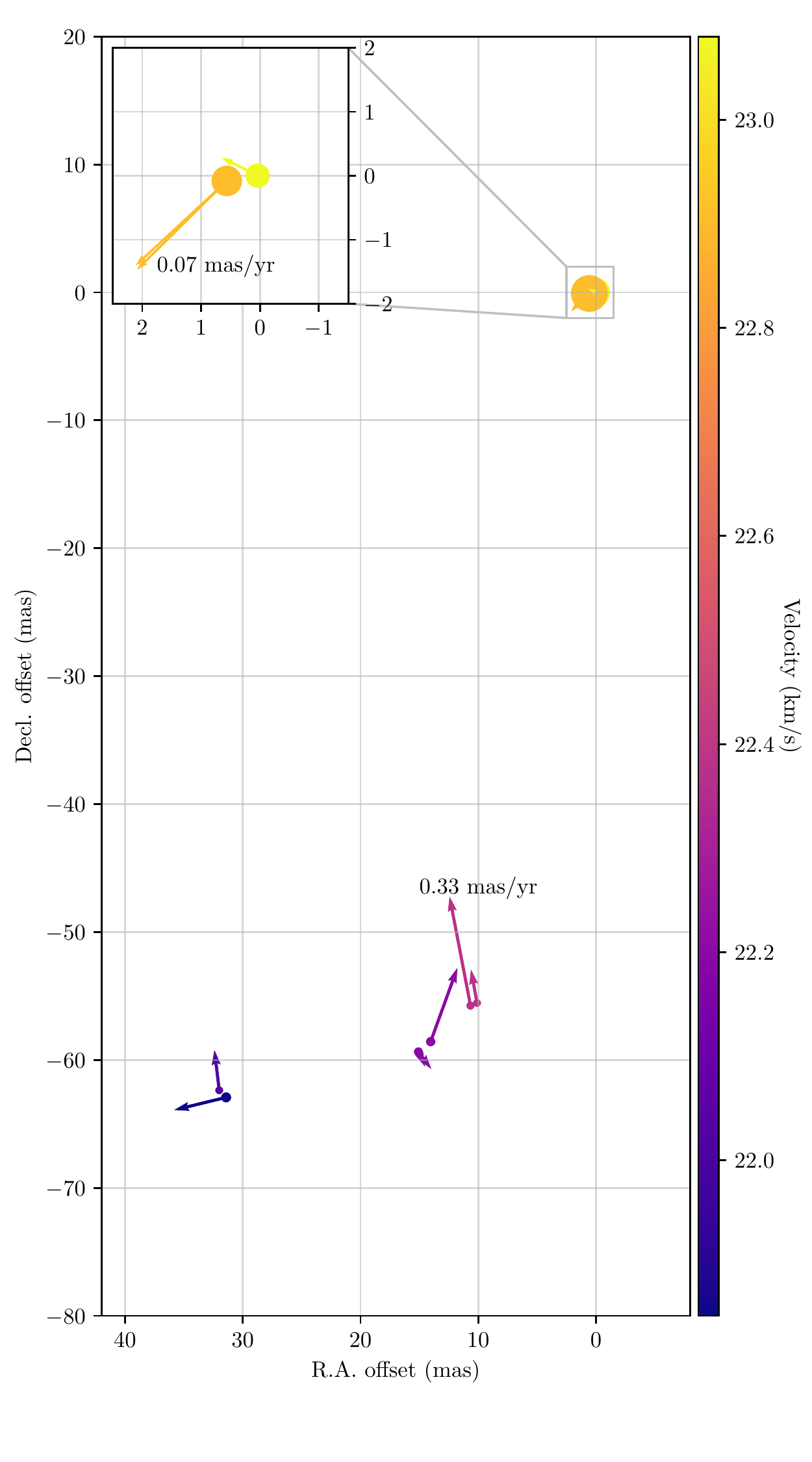}
			\caption{Map of methanol maser spots in G232.62+00.99 and their proper motions relative to the reference spot at (0,0). Features shown were detected in $\ge5$ epochs. \R{ {\bf Top left:} Zoomed-in section around (0,0) shows the main emission region is a blended-double. }}
			\label{fig:g232_spotmap}
	\end{figure}

    \R{The methanol maser emission towards G232.62+00.99 can be grouped into four regions, with the northernmost being the brightest. The regions are distributed in a slightly arched or linear arrangement, and have a velocity gradient running South-East to North-West. The positions and internal motions relative to the reference feature are shown in \Fig{fig:g232_spotmap}. The internal motions of the region relative to the reference maser feature are small, only $\le0.33$~mas/yr or 2.6~\kms\ at the measured distance. This suggests that the internal motion of the reference feature is also small and that the measured proper motion of that feature is representative of the region as a whole. We have inflated the error in the region proper motion to $0.33$~mas/yr (\Tab{tab:parallax}) to account for this. The weakest consistently-detected maser spot was 2~Jy, and the measured maser spot distribution is consistent with that reported by \citet{Fujisawa2014}.}
    
    G232.62+00.99 well-matches the ($l,b$) coordinates of the Local arm as traced by \citet{Reid2019}.   The centroid velocity of the associated 6.7~GHz masers is near 23~\kms, which compares reasonably with 17~\kms\ fitted to Local arm sources nearby in angle.
    
    The center of the Local arm at longitude $232$\degs\ is at
    a distance of 0.83~kpc, and the estimated Gaussian $1\sigma$ width of an arm
    at a Galactocentric radius of 9~kpc is 0.4~kpc \citep{Reid2019}.  At our measured distance of 1.64~kpc, this places G232.62+00.99 at 0.81~kpc (or about $2\sigma$) from the arm center.  Since this source is at one end (at Galactic azimuth $-8$\degs) of sources with measured parallaxes used to trace the Local arm, it could be that the pitch angle fitted over that azimuth range extending to the other end (azimuth $+34$\degs) should be increased slightly.  Interestingly, however, there is a ``bridge'' of gas seen in HI starting at ($l$,$V_{lsr}$) = (232\degs,20~\kms) and connecting to the Perseus arm at (242\degs,70~\kms) \citep[see Fig. 12 of][]{Reid2016}. Possibly, G232.62+00.99 is associated with this bridge.

  \subsection{G323.74--00.26}
     G323.74--00.26 is clearly associated with the Scutum-Centaurus spiral arm,
     since its ($l,b,V_{lsr}$) coordinates of (323\d74,$-0$\d26,$-50$~\kms) compare 
     very well with the arm model of ($323$\degs,$-0$\d01,$-53$~\kms) of \citet{Reid2019}. 
     That model places the center of the arm at this longitude at a distance of 3.2~kpc, which is about 0.45~kpc more distant than our parallax.  At a Galactocentric radius of 6.1~kpc the arm width estimate of \citet{Reid2019} is 0.26~kpc, placing this source 
     $1.7\sigma$ from the center.  However, given that at present very few southern sources have accurate parallax measurements, this is not surprising, since the precise location of the Centaurus arm segment might be fairly uncertain at this time.

	\begin{figure*}[ht]
			\centering
			\includegraphics[width=0.9\textwidth]{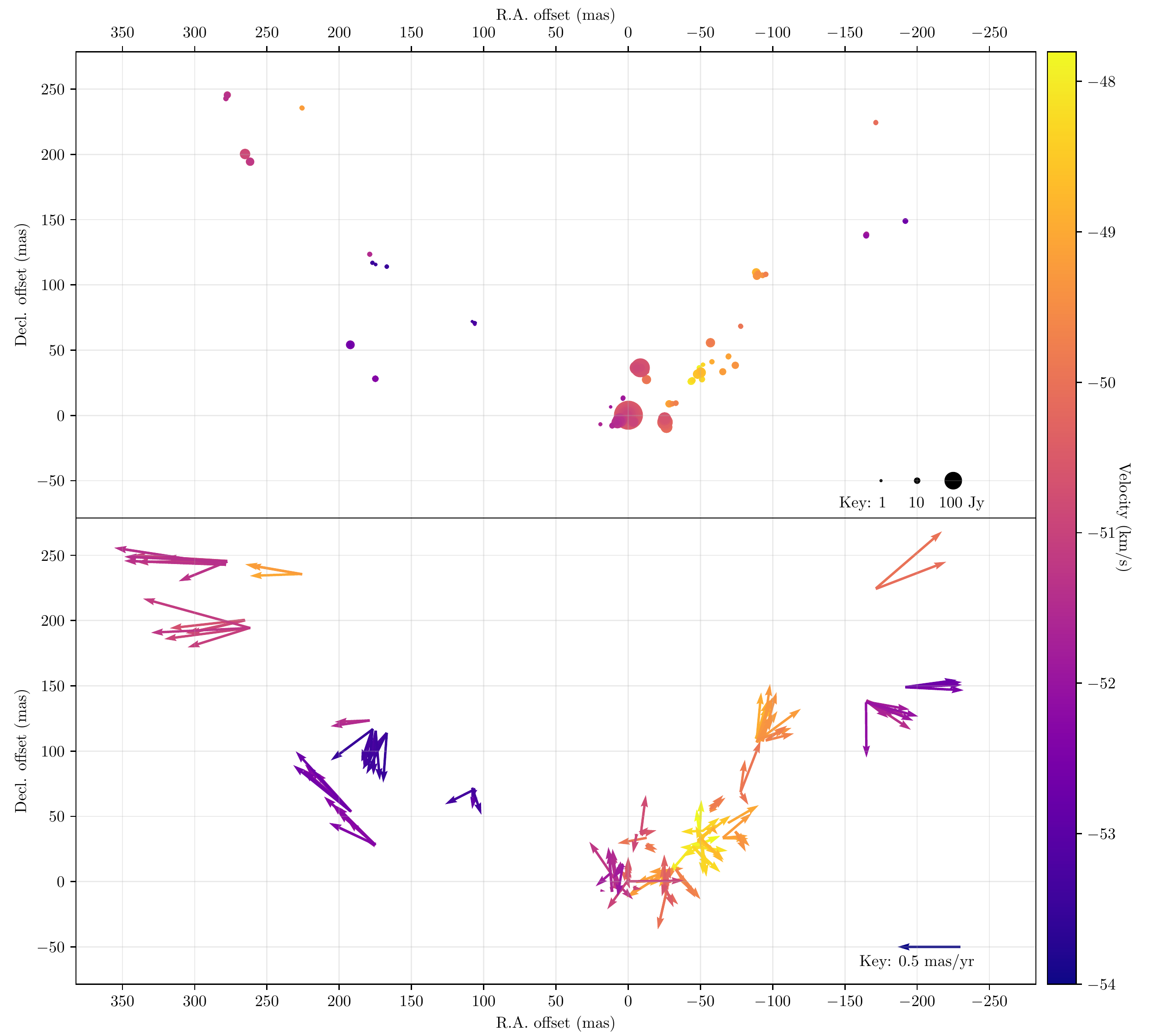}
			\caption{Map \R{and dynamics of maser spots in G323.74--00.26}. \R{{\bf Top:} Positions and flux densities of maser spots. {\bf Bottom}: Internal motions of maser spots.} Features shown were detected in $\ge5$ epochs. }
			\label{fig:g323_spotmap}
	\end{figure*}
    
    The G323.74--00.26 maser emissions arise from quite a large number of spots. \Fig{fig:g323_spotmap} shows the positions of bright spots and their apparent motions over time. We have subtracted the average motion of $\mu_{x_{int}}=0.00\pm0.02$ and $\mu_{y_{int}}=-0.198\pm0.012$~mas~y$^{-1}$ in the R.A. and declination directions respectively. If the distribution of measured spot motion is near-isotropic, this average will reflect the internal motion of the reference feature. Therefore we also subtracted this motion from the measured proper motion of the reference feature measured with respect to the quasars (\Tab{tab:parallax}) to obtain an estimate of the absolute motion of the region giving $\mu_{x_{tot}}=-3.2\pm0.4$ and $\mu_{y_{tot}}=-4.2\pm0.4$~mas~y$^{-1}$. Here we have added an additional uncertainty of 0.4~mas~y$^{-1}$ in quadrature (equivalent to 5~\kms\ at the measured distance) to account for the likelihood that the spots do not have an isotropic velocity distribution. There does not seem to be significant evidence in favor of an edge-on disk structure as has been suggested for this methanol maser \citep[e.g.,][]{Phillips1998}, and the structure and internal motions instead suggest that the maser spot distribution may be part of a bow shock.

    The peculiar (non-circular) motion of G323.74--00.26 about the Galactic Center of mass can be calculated from its measured 6-dimensional phase-space values.  Adopting a $V_{lsr}=50.5\pm5.0$ \kms\
    and the rotation curve of \citet{Reid2019}, we find ($U_p,V_p,W_p$) = ($-5,-1,-8$) \kms, where $U_p$ is toward the Galactic center at the position of the source, $V_p$ is in the direction of Galactic rotation, and $W_P$ is toward the North Galactic Pole.   Uncertainties from measurement error are $\pm5$ \kms\ in each coordinate, so G323.74--00.26 has a very small peculiar motion as is typical for a young high-mass star.
    
    \subsection{Inverse MultiView}
    Comparing the parallax and proper motion results from iMV with a group of calibrators to standard (inverse) phase referencing with a single calibrator  (\Tab{tab:parallax}), we find that iMV is at least a factor of two better in accuracy.
    In \citet{Hyland2022}, per-epoch positional uncertainties of $\pm20~\mu$as were achieved at 8.3~GHz for calibrator separations $<7$\degs\ in both NS and EW directions. Here we report per-epoch positional uncertainties (determined from the error floor values) of $\approx26$~$\mu$as in the EW direction and $\approx50$~$\mu$as in the NS direction.
    
    \begin{figure*}[ht]
        \centering
        \includegraphics[width=0.49\textwidth]{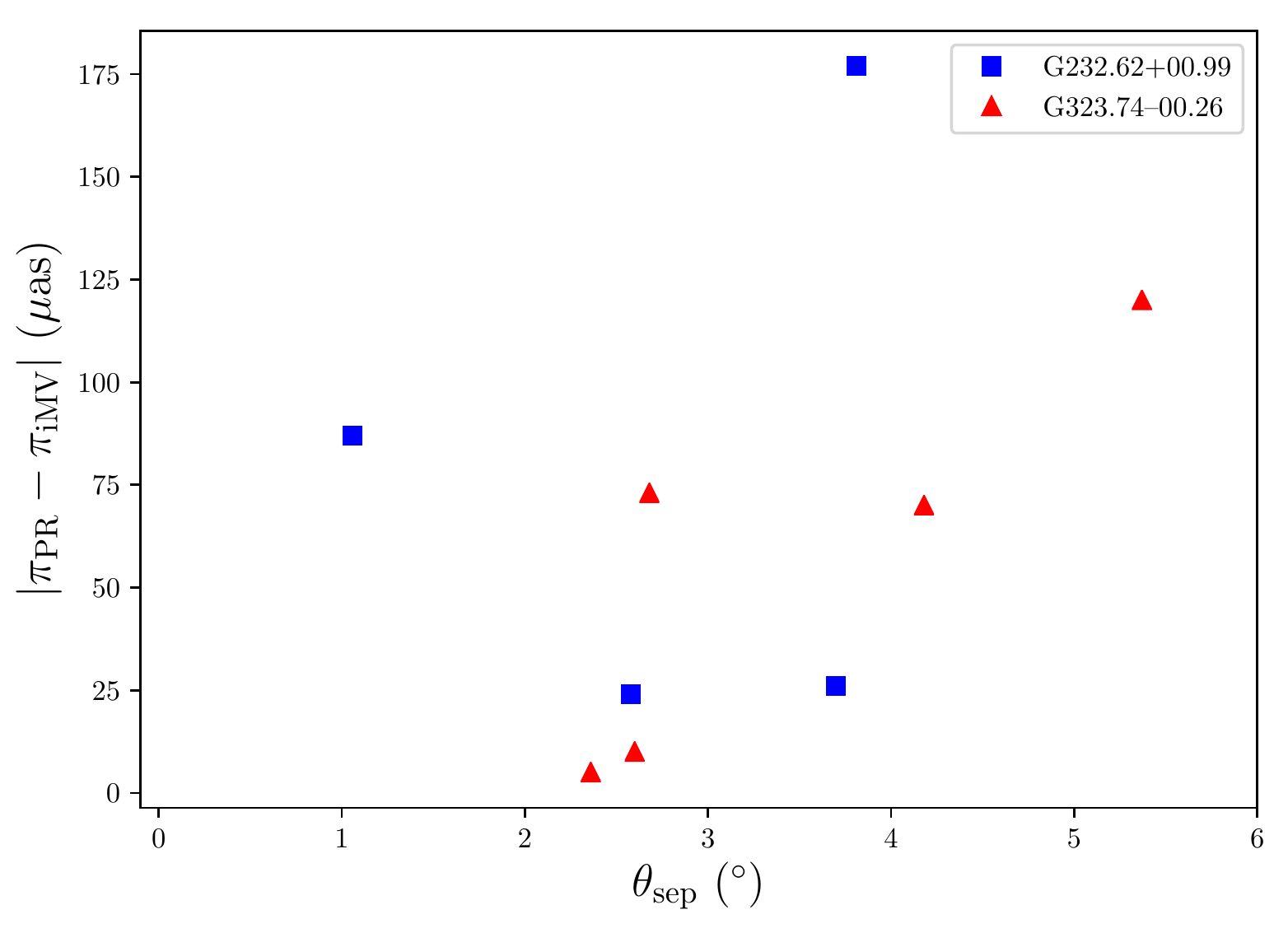}
        \caption{\R{Absolute difference between parallaxes determined with iMV and iPR for each maser.} }
        \label{fig:errorvssep}
    \end{figure*}
    
    \Fig{fig:errorvssep} shows the \R{difference  between the iMV and} iPR parallaxes as a function of total angular separation for both targets. \R{Under the assumption that the iMV parallax is the true parallax, this plot shows the systematic error of iPR parallaxes vs. target-calibrator separation.} The behavior of the G323.74--00.26 iPR parallaxes is exactly as expected, where the error increases with target-calibrator separation. \R{For G232.62+00.99 there is a larger spread and no clear trend. In both cases, all parallaxes based on different quasars have much more uncertain parallax fits.
    
    \R{Comparisons between the relative performance of iMV and iPR is inequitable because the experiments were specifically tailored for iMV. However, in comparison with past 6.7~GHz parallax experiments \citep[e.g.,][]{Krishnan2015,Krishnan2017,Reid2017} is it clear that our additional calibration steps have reduced astrometric uncertainty, and/or removed possible systematic errors that may have been introduced in the parallax fit.}

    It is important to note that for at least one quasar per maser, an iPR parallax was unable to be determined. Once for the most distant calibrator (J1730--141), and the other for an intermediate distance calibrator (J1524--5903, \Tab{tab:parallax}).  In both cases, the quasar in question was not visible in synthesized images at a number of epochs, presumably due to the uncalibrated delay slopes and perhaps additionally in the case of J1524, low flux density. 
    
    For this reason, we stress the use of numerous compact and strong calibrators with good positions, which allow the original maser position to be updated despite the presence of uncalibrated delays (i.e., in step 6 of preliminary reduction). It appears that five calibrators are the ideal number, allowing a balance between spatial and temporal sampling, while four calibrators should be considered the bare minimum. For specific linear arrangements of the target and calibrators, fewer may be acceptable.} 
        
    \section{Concluding Remarks}
    We have used inverse MultiView to measure the parallax and proper motion of two 6.7~GHz class II methanol masers, with results approaching the highest accuracy ever achieved at this frequency.

    As part of the \spirals\ project, we will continue to measure the parallaxes and proper motions for southern hemisphere 6.7~GHz class II methanol masers to fill in the 4th quadrant to better trace the spiral arms of the Milky Way. \R{Based on the current array capabilities and preliminary surveys of compact maser emission on VLBI baselines \citep{Hyland2021}, we estimate there are at least 20 masers that are bright and compact enough to measure parallax with iMV in the current iteration of \spirals. We also aim to explore the application of direct MV \citep{Rioja2017} to measure parallaxes and proper motions to the weaker methanol masers inappropriate for iMV, which may double the number of possible targets.}

\section*{Data and code availability}
	The data underlying this article will be shared on reasonable request to the corresponding author. The scripts and programs used for data reduction and calibration are available at \url{https://github.com/lucasjord/spirals/script}. \R{Data behind Figures 1, 2, 7 and 8 are available at \url{https://github.com/lucasjord/spirals/imultiview_p2}.}

\section*{ACKNOWLEDGEMENTS}
		
\indent This research was supported by the Australian Research Council (ARC) Discovery Grant DP180101061. We want to thank Mr Brett Reid and Mrs Bev Benson for maintaining and organizing the University of Tasmania radio telescopes. We acknowledge the Jawoyn, Paredarerme, and Wiriangu peoples as the traditional owners of the land situating the Katherine, Hobart, and Ceduna telescopes respectively. The Warkworth~30m radio telescope is funded and operated by the Auckland University of Technology; we would like to thank Mr. Lewis Woodburn for the maintenance and management of this facility to enable its involvement in this project. This research has made use of NASA’s Astrophysics Data System Abstract Service. This research made use of \url{MaserDB.net}, an online database of astrophysical masers \citep{maserdb2019}. This research made use of Astropy, a community-developed core Python package for Astronomy \citep{astropy:2013,astropy:2018}.

\pagebreak
\section*{APPENDIX}
The model used to predict the position of the target $(x,y)$ at any given time ($t$) relative to some reference time ($t_0$) due to parallax ($\pi$) and proper motion ($\mu_x, \mu_y$) is as follows:
\begin{equation}
\begin{aligned}
x &= \pi (Y\cos\alpha - X\sin\alpha) + \mu_x(t-t_0) \\
y &= \pi (Z\cos\delta - X\cos\alpha\sin\delta \\
&\quad - Y\sin\alpha\sin\delta) + \mu_y(t-t_0)
\end{aligned}
\label{eq:model_parallax}
\end{equation}
where $\alpha$ and $\delta$ are the R.A. and Dec. of the target, and $X$, $Y$, $Z$ represent the Earth's position relative to the Sun at epoch $t$ (as determined by the NOVAS subroutines; \cite{1989AJ.....97.1197K}). Since the NOVAS routines request $t$ in MJD, the $\mu$ values are returned in mas d$^{-1}$.

		
\bibliographystyle{yahapj}
\bibliography{multiview_paper}

\begin{thebibliography}{}
\providecommand\natexlab[1]{#1}
\providecommand\JournalTitle[1]{#1}

\bibitem[{{Astropy Collaboration} {et~al.}(2013){Astropy Collaboration},
  {Robitaille}, {Tollerud}, {Greenfield}, {Droettboom}, {Bray}, {Aldcroft},
  {Davis}, {Ginsburg}, {Price-Whelan}, {Kerzendorf}, {Conley}, {Crighton},
  {Barbary}, {Muna}, {Ferguson}, {Grollier}, {Parikh}, {Nair}, {Unther},
  {Deil}, {Woillez}, {Conseil}, {Kramer}, {Turner}, {Singer}, {Fox}, {Weaver},
  {Zabalza}, {Edwards}, {Azalee Bostroem}, {Burke}, {Casey}, {Crawford},
  {Dencheva}, {Ely}, {Jenness}, {Labrie}, {Lim}, {Pierfederici}, {Pontzen},
  {Ptak}, {Refsdal}, {Servillat}, \& {Streicher}}]{astropy:2013}
{Astropy Collaboration}, {Robitaille}, T.~P., {Tollerud}, E.~J., {et~al.} 2013,
  \href{http://dx.doi.org/10.1051/0004-6361/201322068}{\JournalTitle{\aap},
  558, A33}

\bibitem[{{Astropy Collaboration} {et~al.}(2018){Astropy Collaboration},
  {Price-Whelan}, {Sip{\H{o}}cz}, {G{\"u}nther}, {Lim}, {Crawford}, {Conseil},
  {Shupe}, {Craig}, {Dencheva}, {Ginsburg}, {Vand erPlas}, {Bradley},
  {P{\'e}rez-Su{\'a}rez}, {de Val-Borro}, {Aldcroft}, {Cruz}, {Robitaille},
  {Tollerud}, {Ardelean}, {Babej}, {Bach}, {Bachetti}, {Bakanov}, {Bamford},
  {Barentsen}, {Barmby}, {Baumbach}, {Berry}, {Biscani}, {Boquien}, {Bostroem},
  {Bouma}, {Brammer}, {Bray}, {Breytenbach}, {Buddelmeijer}, {Burke},
  {Calderone}, {Cano Rodr{\'\i}guez}, {Cara}, {Cardoso}, {Cheedella}, {Copin},
  {Corrales}, {Crichton}, {D'Avella}, {Deil}, {Depagne}, {Dietrich}, {Donath},
  {Droettboom}, {Earl}, {Erben}, {Fabbro}, {Ferreira}, {Finethy}, {Fox},
  {Garrison}, {Gibbons}, {Goldstein}, {Gommers}, {Greco}, {Greenfield},
  {Groener}, {Grollier}, {Hagen}, {Hirst}, {Homeier}, {Horton}, {Hosseinzadeh},
  {Hu}, {Hunkeler}, {Ivezi{\'c}}, {Jain}, {Jenness}, {Kanarek}, {Kendrew},
  {Kern}, {Kerzendorf}, {Khvalko}, {King}, {Kirkby}, {Kulkarni}, {Kumar},
  {Lee}, {Lenz}, {Littlefair}, {Ma}, {Macleod}, {Mastropietro}, {McCully},
  {Montagnac}, {Morris}, {Mueller}, {Mumford}, {Muna}, {Murphy}, {Nelson},
  {Nguyen}, {Ninan}, {N{\"o}the}, {Ogaz}, {Oh}, {Parejko}, {Parley}, {Pascual},
  {Patil}, {Patil}, {Plunkett}, {Prochaska}, {Rastogi}, {Reddy Janga},
  {Sabater}, {Sakurikar}, {Seifert}, {Sherbert}, {Sherwood-Taylor}, {Shih},
  {Sick}, {Silbiger}, {Singanamalla}, {Singer}, {Sladen}, {Sooley},
  {Sornarajah}, {Streicher}, {Teuben}, {Thomas}, {Tremblay}, {Turner},
  {Terr{\'o}n}, {van Kerkwijk}, {de la Vega}, {Watkins}, {Weaver}, {Whitmore},
  {Woillez}, {Zabalza}, \& {Astropy Contributors}}]{astropy:2018}
{Astropy Collaboration}, {Price-Whelan}, A.~M., {Sip{\H{o}}cz}, B.~M., {et~al.}
  2018, \href{http://dx.doi.org/10.3847/1538-3881/aabc4f}{\JournalTitle{\aj},
  156, 123}

\bibitem[{{Breen} {et~al.}(2013){Breen}, {Ellingsen}, {Contreras}, {Green},
  {Caswell}, {Stevens}, {Dawson}, \& {Voronkov}}]{Breen2013}
{Breen}, S.~L., {Ellingsen}, S.~P., {Contreras}, Y., {et~al.} 2013,
  \href{http://dx.doi.org/10.1093/mnras/stt1315}{\JournalTitle{\mnras}, 435,
  524}

\bibitem[{{Breen} {et~al.}(2015){Breen}, {Fuller}, {Caswell}, {Green},
  {Avison}, {Ellingsen}, {Gray}, {Pestalozzi}, {Quinn}, {Richards}, {Thompson},
  \& {Voronkov}}]{Breen2015}
{Breen}, S.~L., {Fuller}, G.~A., {Caswell}, J.~L., {et~al.} 2015,
  \href{http://dx.doi.org/10.1093/mnras/stv847}{\JournalTitle{\mnras}, 450,
  4109}

\bibitem[{{Brunthaler} {et~al.}(2011){Brunthaler}, {Reid}, {Menten}, {Zheng},
  {Bartkiewicz}, {Choi}, {Dame}, {Hachisuka}, {Immer}, {Moellenbrock},
  {Moscadelli}, {Rygl}, {Sanna}, {Sato}, {Wu}, {Xu}, \&
  {Zhang}}]{Brunthaler2011}
{Brunthaler}, A., {Reid}, M.~J., {Menten}, K.~M., {et~al.} 2011,
  \href{http://dx.doi.org/10.1002/asna.201111560}{\JournalTitle{Astronomische
  Nachrichten}, 332, 461}

\bibitem[{{Caswell} {et~al.}(2010){Caswell}, {Fuller}, {Green}, {Avison},
  {Breen}, {Brooks}, {Burton}, {Chrysostomou}, {Cox}, {Diamond}, {Ellingsen},
  {Gray}, {Hoare}, {Masheder}, {McClure-Griffiths}, {Pestalozzi}, {Phillips},
  {Quinn}, {Thompson}, {Voronkov}, {Walsh}, {Ward-Thompson}, {Wong-McSweeney},
  {Yates}, \& {Cohen}}]{Caswell2010}
{Caswell}, J.~L., {Fuller}, G.~A., {Green}, J.~A., {et~al.} 2010,
  \href{http://dx.doi.org/10.1111/j.1365-2966.2010.16339.x}{\JournalTitle{\mnras},
  404, 1029}

\bibitem[{{Caswell} {et~al.}(2011){Caswell}, {Fuller}, {Green}, {Avison},
  {Breen}, {Ellingsen}, {Gray}, {Pestalozzi}, {Quinn}, {Thompson}, \&
  {Voronkov}}]{Caswell2011}
---. 2011,
  \href{http://dx.doi.org/10.1111/j.1365-2966.2011.19383.x}{\JournalTitle{\mnras},
  417, 1964}

\bibitem[{{Deller} {et~al.}(2011){Deller}, {Brisken}, {Phillips}, {Morgan},
  {Alef}, {Cappallo}, {Middelberg}, {Romney}, {Rottmann}, {Tingay}, \&
  {Wayth}}]{Deller2011}
{Deller}, A.~T., {Brisken}, W.~F., {Phillips}, C.~J., {et~al.} 2011,
  \href{http://dx.doi.org/10.1086/658907}{\JournalTitle{\pasp}, 123, 275}

\bibitem[{{Dodson}(2008)}]{Dodson2008}
{Dodson}, R. 2008,
  \href{http://dx.doi.org/10.1051/0004-6361:20078670}{\JournalTitle{\aap}, 480,
  767}

\bibitem[{{Dodson} \& {Rioja}(2022)}]{beamwaveguide}
{Dodson}, R., \& {Rioja}, M.~J. 2022, in eLBA memo 14

\bibitem[{{Dubout-Crillon}(1976)}]{DuboutCrillon1976}
{Dubout-Crillon}, R. 1976, \JournalTitle{\aaps}, 25, 25

\bibitem[{{Ellingsen}(2007)}]{Ellingsen2007}
{Ellingsen}, S.~P. 2007,
  \href{http://dx.doi.org/10.1111/j.1365-2966.2007.11615.x}{\JournalTitle{\mnras},
  377, 571}

\bibitem[{{Fujisawa} {et~al.}(2014){Fujisawa}, {Sugiyama}, {Motogi},
  {Hachisuka}, {Yonekura}, {Sawada-Satoh}, {Matsumoto}, {Sorai}, {Momose},
  {Saito}, {Takaba}, {Ogawa}, {Kimura}, {Niinuma}, {Hirano}, {Omodaka},
  {Kobayashi}, {Kawaguchi}, {Shibata}, {Honma}, {Hirota}, {Murata}, {Doi},
  {Mochizuki}, {Shen}, {Chen}, {Xia}, {Li}, \& {Kim}}]{Fujisawa2014}
{Fujisawa}, K., {Sugiyama}, K., {Motogi}, K., {et~al.} 2014,
  \href{http://dx.doi.org/10.1093/pasj/psu015}{\JournalTitle{\pasj}, 66, 31}

\bibitem[{{Goedhart} {et~al.}(2004){Goedhart}, {Gaylard}, \& {van der
  Walt}}]{Goedhart2004}
{Goedhart}, S., {Gaylard}, M.~J., \& {van der Walt}, D.~J. 2004,
  \href{http://dx.doi.org/10.1111/j.1365-2966.2004.08340.x}{\JournalTitle{\mnras},
  355, 553}

\bibitem[{{Green} {et~al.}(2012){Green}, {Caswell}, {Fuller}, {Avison},
  {Breen}, {Ellingsen}, {Gray}, {Pestalozzi}, {Quinn}, {Thompson}, \&
  {Voronkov}}]{Green2012}
{Green}, J.~A., {Caswell}, J.~L., {Fuller}, G.~A., {et~al.} 2012,
  \href{http://dx.doi.org/10.1111/j.1365-2966.2011.20229.x}{\JournalTitle{\mnras},
  420, 3108}

\bibitem[{{Greisen}(1990)}]{Greisen1990}
{Greisen}, E.~W. 1990, in Acquisition, Processing and Archiving of Astronomical
  Images, 125

\bibitem[{{Greisen}(2003)}]{Greisen2003}
{Greisen}, E.~W. 2003, \href{http://dx.doi.org/10.1007/0-306-48080-8_7}{in
  Astrophysics and Space Science Library, Vol. 285, Information Handling in
  Astronomy - Historical Vistas, ed. A.~{Heck}}, 109

\bibitem[{{Honma} {et~al.}(2007){Honma}, {Bushimata}, {Choi}, {Hirota}, {Imai},
  {Iwadate}, {Jike}, {Kameya}, {Kamohara}, {Kan-Ya}, {Kawaguchi}, {Kijima},
  {Kobayashi}, {Kuji}, {Kurayama}, {Manabe}, {Miyaji}, {Nagayama}, {Nakagawa},
  {Oh}, {Omodaka}, {Oyama}, {Sakai}, {Sato}, {Sasao}, {Shibata}, {Shintani},
  {Suda}, {Tamura}, {Tsushima}, \& {Yamashita Kazuyoshi}}]{Honma2007}
{Honma}, M., {Bushimata}, T., {Choi}, Y.~K., {et~al.} 2007,
  \href{http://dx.doi.org/10.1093/pasj/59.5.889}{\JournalTitle{\pasj}, 59, 889}

\bibitem[{{Hyland}(2021)}]{Hyland2021}
{Hyland}, L.~J. 2021, PhD thesis, School of Natural Sciences

\bibitem[{{Hyland} {et~al.}(2022){Hyland}, {Reid}, {Ellingsen}, {Rioja},
  {Dodson}, {Orosz}, {Masson}, \& {McCallum}}]{Hyland2022}
{Hyland}, L.~J., {Reid}, M.~J., {Ellingsen}, S.~P., {et~al.} 2022,
  \href{http://dx.doi.org/10.3847/1538-4357/ac6d5b}{\JournalTitle{\apj}, 932,
  52}

\bibitem[{{Kaplan} {et~al.}(1989){Kaplan}, {Hughes}, {Seidelmann}, {Smith}, \&
  {Yallop}}]{1989AJ.....97.1197K}
{Kaplan}, G.~H., {Hughes}, J.~A., {Seidelmann}, P.~K., {Smith}, C.~A., \&
  {Yallop}, B.~D. 1989,
  \href{http://dx.doi.org/10.1086/115063}{\JournalTitle{\aj}, 97, 1197}

\bibitem[{{Kettenis} {et~al.}(2006){Kettenis}, {van Langevelde}, {Reynolds}, \&
  {Cotton}}]{Kettenis2006}
{Kettenis}, M., {van Langevelde}, H.~J., {Reynolds}, C., \& {Cotton}, B. 2006,
  in Astronomical Society of the Pacific Conference Series, Vol. 351,
  Astronomical Data Analysis Software and Systems XV, ed. C.~{Gabriel},
  C.~{Arviset}, D.~{Ponz}, \& S.~{Enrique}, 497

\bibitem[{{Krishnan} {et~al.}(2017){Krishnan}, {Ellingsen}, {Reid}, {Bignall},
  {McCallum}, {Phillips}, {Reynolds}, \& {Stevens}}]{Krishnan2017}
{Krishnan}, V., {Ellingsen}, S.~P., {Reid}, M.~J., {et~al.} 2017,
  \href{http://dx.doi.org/10.1093/mnras/stw2850}{\JournalTitle{\mnras}, 465,
  1095}

\bibitem[{{Krishnan} {et~al.}(2015){Krishnan}, {Ellingsen}, {Reid},
  {Brunthaler}, {Sanna}, {McCallum}, {Reynolds}, {Bignall}, {Phillips},
  {Dodson}, {Rioja}, {Caswell}, {Chen}, {Dawson}, {Fujisawa}, {Goedhart},
  {Green}, {Hachisuka}, {Honma}, {Menten}, {Shen}, {Voronkov}, {Walsh}, {Xu},
  {Zhang}, \& {Zheng}}]{Krishnan2015}
---. 2015,
  \href{http://dx.doi.org/10.1088/0004-637X/805/2/129}{\JournalTitle{\apj},
  805, 129}

\bibitem[{{Ladeyschikov} {et~al.}(2019){Ladeyschikov}, {Bayandina}, \&
  {Sobolev}}]{maserdb2019}
{Ladeyschikov}, D.~A., {Bayandina}, O.~S., \& {Sobolev}, A.~M. 2019,
  \href{http://dx.doi.org/10.3847/1538-3881/ab4b4c}{\JournalTitle{\aj}, 158,
  233}

\bibitem[{{Lovell} {et~al.}(2013){Lovell}, {McCallum}, {Reid}, {McCulloch},
  {Baynes}, {Dickey}, {Shabala}, {Watson}, {Titov}, {Ruddick}, {Twilley},
  {Reynolds}, {Tingay}, {Shield}, {Adada}, {Ellingsen}, {Morgan}, \&
  {Bignall}}]{Lovell2013}
{Lovell}, J.~E.~J., {McCallum}, J.~N., {Reid}, P.~B., {et~al.} 2013,
  \href{http://dx.doi.org/10.1007/s00190-013-0626-3}{\JournalTitle{Journal of
  Geodesy}, 87, 527}

\bibitem[{{MacLeod} {et~al.}(1992){MacLeod}, {Gaylard}, \&
  {Nicolson}}]{MacLeod1992}
{MacLeod}, G.~C., {Gaylard}, M.~J., \& {Nicolson}, G.~D. 1992,
  \href{http://dx.doi.org/10.1093/mnras/254.1.1P}{\JournalTitle{\mnras}, 254,
  1P}

\bibitem[{{McCulloch} {et~al.}(2005){McCulloch}, {Ellingsen}, {Jauncey},
  {Carter}, {Cim{\`o}}, {Lovell}, \& {Dodson}}]{McCulloch2005}
{McCulloch}, P.~M., {Ellingsen}, S.~P., {Jauncey}, D.~L., {et~al.} 2005,
  \href{http://dx.doi.org/10.1086/428374}{\JournalTitle{\aj}, 129, 2034}

\bibitem[{{Menten}(1991)}]{Menten1991}
{Menten}, K.~M. 1991,
  \href{http://dx.doi.org/10.1086/186177}{\JournalTitle{\apjl}, 380, L75}

\bibitem[{{Minier} {et~al.}(2001){Minier}, {Conway}, \& {Booth}}]{Minier2001}
{Minier}, V., {Conway}, J.~E., \& {Booth}, R.~S. 2001,
  \href{http://dx.doi.org/10.1051/0004-6361:20010124}{\JournalTitle{\aap}, 369,
  278}

\bibitem[{{Norris} {et~al.}(1993){Norris}, {Whiteoak}, {Caswell}, {Wieringa},
  \& {Gough}}]{Norris1993}
{Norris}, R.~P., {Whiteoak}, J.~B., {Caswell}, J.~L., {Wieringa}, M.~H., \&
  {Gough}, R.~G. 1993,
  \href{http://dx.doi.org/10.1086/172914}{\JournalTitle{\apj}, 412, 222}

\bibitem[{{Norris} {et~al.}(1998){Norris}, {Byleveld}, {Diamond}, {Ellingsen},
  {Ferris}, {Gough}, {Kesteven}, {McCulloch}, {Phillips}, {Reynolds},
  {Tzioumis}, {Takahashi}, {Troup}, \& {Wellington}}]{Norris1998}
{Norris}, R.~P., {Byleveld}, S.~E., {Diamond}, P.~J., {et~al.} 1998,
  \href{http://dx.doi.org/10.1086/306373}{\JournalTitle{\apj}, 508, 275}

\bibitem[{{Petrov} {et~al.}(2019){Petrov}, {de Witt}, {Sadler}, {Phillips}, \&
  {Horiuchi}}]{Petrov2019}
{Petrov}, L., {de Witt}, A., {Sadler}, E.~M., {Phillips}, C., \& {Horiuchi}, S.
  2019, \href{http://dx.doi.org/10.1093/mnras/stz242}{\JournalTitle{\mnras},
  485, 88}

\bibitem[{{Petrov} {et~al.}(2015){Petrov}, {Natusch}, {Weston}, {McCallum},
  {Ellingsen}, \& {Gulyaev}}]{Petrov2015}
{Petrov}, L., {Natusch}, T., {Weston}, S., {et~al.} 2015,
  \href{http://dx.doi.org/10.1086/681965}{\JournalTitle{\pasp}, 127, 516}

\bibitem[{{Phillips} {et~al.}(1998){Phillips}, {Norris}, {Ellingsen}, \&
  {McCulloch}}]{Phillips1998}
{Phillips}, C.~J., {Norris}, R.~P., {Ellingsen}, S.~P., \& {McCulloch}, P.~M.
  1998,
  \href{http://dx.doi.org/10.1046/j.1365-8711.1998.01979.x}{\JournalTitle{\mnras},
  300, 1131}

\bibitem[{{Reid} {et~al.}(2016){Reid}, {Dame}, {Menten}, \&
  {Brunthaler}}]{Reid2016}
{Reid}, M.~J., {Dame}, T.~M., {Menten}, K.~M., \& {Brunthaler}, A. 2016,
  \href{http://dx.doi.org/10.3847/0004-637X/823/2/77}{\JournalTitle{\apj}, 823,
  77}

\bibitem[{{Reid} \& {Honma}(2014)}]{ReidHonma2014}
{Reid}, M.~J., \& {Honma}, M. 2014,
  \href{http://dx.doi.org/10.1146/annurev-astro-081913-040006}{\JournalTitle{\araa},
  52, 339}

\bibitem[{{Reid} {et~al.}(2009{\natexlab{a}}){Reid}, {Menten}, {Brunthaler},
  {Zheng}, {Moscadelli}, \& {Xu}}]{Reid2009_mas}
{Reid}, M.~J., {Menten}, K.~M., {Brunthaler}, A., {et~al.} 2009{\natexlab{a}},
  \href{http://dx.doi.org/10.1088/0004-637X/693/1/397}{\JournalTitle{\apj},
  693, 397}

\bibitem[{{Reid} {et~al.}(2009{\natexlab{b}}){Reid}, {Menten}, {Zheng},
  {Brunthaler}, {Moscadelli}, {Xu}, {Zhang}, {Sato}, {Honma}, {Hirota},
  {Hachisuka}, {Choi}, {Moellenbrock}, \& {Bartkiewicz}}]{Reid2009_gal}
{Reid}, M.~J., {Menten}, K.~M., {Zheng}, X.~W., {et~al.} 2009{\natexlab{b}},
  \href{http://dx.doi.org/10.1088/0004-637X/700/1/137}{\JournalTitle{\apj},
  700, 137}

\bibitem[{{Reid} {et~al.}(2014){Reid}, {Menten}, {Brunthaler}, {Zheng}, {Dame},
  {Xu}, {Wu}, {Zhang}, {Sanna}, {Sato}, {Hachisuka}, {Choi}, {Immer},
  {Moscadelli}, {Rygl}, \& {Bartkiewicz}}]{Reid2014}
{Reid}, M.~J., {Menten}, K.~M., {Brunthaler}, A., {et~al.} 2014,
  \href{http://dx.doi.org/10.1088/0004-637X/783/2/130}{\JournalTitle{\apj},
  783, 130}

\bibitem[{{Reid} {et~al.}(2017){Reid}, {Brunthaler}, {Menten}, {Sanna}, {Xu},
  {Li}, {Wu}, {Hu}, {Zheng}, {Zhang}, {Immer}, {Rygl}, {Moscadelli}, {Sakai},
  {Bartkiewicz}, \& {Choi}}]{Reid2017}
{Reid}, M.~J., {Brunthaler}, A., {Menten}, K.~M., {et~al.} 2017,
  \href{http://dx.doi.org/10.3847/1538-3881/aa7850}{\JournalTitle{\aj}, 154,
  63}

\bibitem[{{Reid} {et~al.}(2019){Reid}, {Menten}, {Brunthaler}, {Zheng}, {Dame},
  {Xu}, {Li}, {Sakai}, {Wu}, {Immer}, {Zhang}, {Sanna}, {Moscadelli}, {Rygl},
  {Bartkiewicz}, {Hu}, {Quiroga-Nu{\~n}ez}, \& {van Langevelde}}]{Reid2019}
{Reid}, M.~J., {Menten}, K.~M., {Brunthaler}, A., {et~al.} 2019,
  \href{http://dx.doi.org/10.3847/1538-4357/ab4a11}{\JournalTitle{\apj}, 885,
  131}

\bibitem[{{Rioja} \& {Dodson}(2020)}]{RiojaDodson2020}
{Rioja}, M.~J., \& {Dodson}, R. 2020,
  \href{http://dx.doi.org/10.1007/s00159-020-00126-z}{\JournalTitle{\aapr}, 28,
  6}

\bibitem[{{Rioja} {et~al.}(2017){Rioja}, {Dodson}, {Orosz}, {Imai}, \&
  {Frey}}]{Rioja2017}
{Rioja}, M.~J., {Dodson}, R., {Orosz}, G., {Imai}, H., \& {Frey}, S. 2017,
  \href{http://dx.doi.org/10.3847/1538-3881/153/3/105}{\JournalTitle{\aj}, 153,
  105}

\bibitem[{{Surcis} {et~al.}(2022){Surcis}, {Vlemmings}, {van Langevelde},
  {Hutawarakorn Kramer}, \& {Bartkiewicz}}]{Surcis2022}
{Surcis}, G., {Vlemmings}, W.~H.~T., {van Langevelde}, H.~J., {Hutawarakorn
  Kramer}, B., \& {Bartkiewicz}, A. 2022,
  \href{http://dx.doi.org/10.1051/0004-6361/202142125}{\JournalTitle{\aap},
  658, A78}

\bibitem[{{VERA Collaboration} {et~al.}(2020){VERA Collaboration}, {Hirota},
  {Nagayama}, {Honma}, {Adachi}, {Burns}, {Chibueze}, {Choi}, {Hachisuka},
  {Hada}, {Hagiwara}, {Hamada}, {Hand a}, {Hashimoto}, {Hirano}, {Hirata},
  {Ichikawa}, {Imai}, {Inenaga}, {Ishikawa}, {Jike}, {Kameya}, {Kaseda}, {Kim},
  {Kim}, {Kim}, {Kobayashi}, {Kono}, {Kurayama}, {Matsuno}, {Morita}, {Motogi},
  {Murase}, {Nakagawa}, {Nakanishi}, {Niinuma}, {Nishi}, {Oh}, {Omodaka},
  {Oyadomari}, {Oyama}, {Sakai}, {Sakai}, {Sawada-Satoh}, {Shibata},
  {Shizugami}, {Sudo}, {Sugiyama}, {Sunada}, {Suzuki}, {Takahashi}, {Tamura},
  {Tazaki}, {Ueno}, {Uno}, {Urago}, {Wada}, {Wu}, {Yamashita}, {Yamashita},
  {Yamauchi}, \& {Yuda}}]{vera2020}
{VERA Collaboration}, {Hirota}, T., {Nagayama}, T., {et~al.} 2020,
  \href{http://dx.doi.org/10.1093/pasj/psaa018}{\JournalTitle{\pasj}},
  \href{http://arxiv.org/abs/2002.03089}{{\sffamily arXiv:2002.03089
  [astro-ph.GA]}}

\bibitem[{{Vlemmings} {et~al.}(2011){Vlemmings}, {Torres}, \&
  {Dodson}}]{Vlemmings2011}
{Vlemmings}, W.~H.~T., {Torres}, R.~M., \& {Dodson}, R. 2011,
  \href{http://dx.doi.org/10.1051/0004-6361/201116648}{\JournalTitle{\aap},
  529, A95}

\bibitem[{{Walsh} {et~al.}(2002){Walsh}, {Lee}, \& {Burton}}]{Walsh2002}
{Walsh}, A.~J., {Lee}, J.~K., \& {Burton}, M.~G. 2002,
  \href{http://dx.doi.org/10.1046/j.1365-8711.2002.05006.x}{\JournalTitle{\mnras},
  329, 475}

\bibitem[{{Woodburn} {et~al.}(2015){Woodburn}, {Natusch}, {Weston},
  {Thomasson}, {Godwin}, {Granet}, \& {Gulyaev}}]{Woodburn2015}
{Woodburn}, L., {Natusch}, T., {Weston}, S., {et~al.} 2015,
  \href{http://dx.doi.org/10.1017/pasa.2015.13}{\JournalTitle{\pasa}, 32, 17}

\end{thebibliography}

\end{document}